\documentclass[journal]{IEEEtran}
\usepackage{color}
\usepackage{ifpdf}

\usepackage{cite}

\usepackage{array}	
\usepackage{float}
\usepackage{multirow}

\usepackage{amsmath}
\usepackage{graphicx}

\begin{document}
%
\title{ISAC in 3GPP: Evolution Toward 6G}
%
%
%

\author{Neeraj Varshney,~\IEEEmembership{Senior Member,~IEEE}
\thanks{Neeraj Varshney is with the Communications Technology Laboratory,
National Institute of Standards and Technology (NIST), Gaithersburg,
MD 20899 USA, and also with Prometheus Computing LLC, Cullowhee,
NC 28723 USA (e-mail: neeraj.varshney@nist.gov)}
}

\markboth{submitted for possible publication in IEEE Internet of Things Journal}%
{Varshney \MakeLowercase{\textit{et al.}}: ISAC in 3GPP: Evolution Toward 6G}

\maketitle

\begin{abstract}
Integrated sensing and communication (ISAC) is emerging as an important direction in the Third Generation Partnership Project (3GPP) evolution toward 6G because it allows cellular networks to provide environmental awareness in addition to connectivity. This paper surveys the current 3GPP trajectory from Release~19 feasibility studies to Release~20 radio, protocol, and architecture studies, while distinguishing established requirements, ongoing study assumptions, and possible forward directions. The survey covers service requirements, sensing topologies, channel model evolution beyond 3GPP Technical Report (TR)~38.901, Radio Access Network Working Group~1 (RAN1) physical layer design, Radio Access Network Working Groups~2 and~3 (RAN2 and RAN3) system implications, and the role of sensing-assisted communication. It also synthesizes the main unresolved issues in waveform and reference signal design, multi-node coordination, sensing data reporting, service exposure, privacy, and implementation constraints. By connecting service-level motivations to physical layer, protocol, and architecture implications, the paper provides a standards-centric reading of how 3GPP may evolve toward practical 6G ISAC support.
\end{abstract}


%
\IEEEpeerreviewmaketitle
\vspace{-0.35cm}
\section{Introduction}
Integrated sensing and communication (ISAC) represents a fundamental paradigm shift in wireless system design, where sensing and communication functionalities are deeply integrated to achieve mutual benefits through shared spectrum, hardware, and signal processing resources \cite{rp234069}. As 6G networks evolve beyond the communication-centric architecture of 5G, ISAC emerges as a key enabling technology for realizing the Internet-of-Everything (IoE) vision, where networks not only connect devices but also perceive and understand their physical environment \cite{tr38765}. This Network as a Sensor concept allows cellular infrastructure to provide environmental awareness as a native service, facilitating applications from autonomous transit to gesture-based human computer interaction \cite{tr22837}. In the cellular context, however, the significance of ISAC is determined not only by sensing capability, but also by whether sensing functions can be integrated into interoperable radio procedures, spectrum management, and service exposure mechanisms.

Third Generation Partnership Project (3GPP) work on ISAC has therefore moved beyond isolated feasibility arguments and now spans service requirements, radio evaluation assumptions, protocol implications, and exposure of sensing capabilities \cite{tr38914,iturm2160}. Unlike purely theoretical studies, 3GPP standardization determines how sensing can coexist with incumbent communication procedures, spectrum constraints, and multi-vendor deployment realities. The current trajectory from New Radio (NR) sensing foundations in Release~19 to broader Release~20 studies thus provides a practical basis for assessing how future 6G systems may incorporate ISAC under interoperable assumptions \cite{tr38765}.
\vspace{-0.35cm}
\subsection{Motivation and Background}

The motivation for ISAC in 6G is not simply higher data traffic, but the opportunity to reuse cellular spectrum, infrastructure, synchronization, antenna arrays, and signal processing resources for environmental awareness. Conventional separated radar and communication deployments can duplicate spectrum use and hardware while keeping sensing outside cellular control, coordination, and service exposure. ISAC instead allows cellular nodes to support communication and device-free sensing within a common radio framework, enabling target detection, tracking, environmental mapping, mobility assistance, industrial automation, and context-aware network optimization \cite{tr38765}. Higher-frequency operation in Frequency Range~3 (FR3), millimeter-wave, and sub-terahertz (sub-THz) bands can further improve delay and angular resolution, but practical sensing performance depends on reference-signal design, beam management, reporting granularity, interference control, and architecture-level coordination. This is where the standards perspective becomes essential, since these dimensions are governed by interoperable procedures rather than by waveform theory alone \cite{tr38765}.

From a standardization perspective, 3GPP involvement ensures that ISAC evolves within the cellular ecosystem rather than as a parallel research track \cite{tr22870}. The sequence from Release~19 sensing foundations \cite{rp234069} to Release~20 studies \cite{tr38765,tr38914} provides a structured path toward anticipated normative work in later 6G phases, while the timing and scope of such work remain subject to ongoing study outcomes \cite{iturm2160}. This standardization effort is complemented by external organizations including the International Telecommunication Union (ITU), which identifies ISAC as an International Mobile Telecommunications (IMT)~2030 usage scenario, and the European Telecommunications Standards Institute (ETSI) Industry Specification Group (ISG) ISAC, which contributes pre-standardization inputs on use cases, key performance indicators (KPIs), architectures, and channel assumptions \cite{iturm2160,etsiisgisac}. This broader institutional setting motivates a survey that follows the standards chain from service requirements to radio design and system architecture rather than discussing each layer in isolation.
\vspace{-0.35cm}
\subsection{Scope and Contributions}

\begin{table*}[!ht]
\caption{Survey Scope and 3GPP Reading Map}
\label{tab:intro_scope_map}
\centering
{
\footnotesize
\setlength{\tabcolsep}{4.0pt}
\renewcommand{\arraystretch}{1.05}
\begin{tabular}{|>{\raggedright\arraybackslash}m{3.1cm}|
                >{\raggedright\arraybackslash}m{7.6cm}|
                >{\centering\arraybackslash}m{2.8cm}|
                >{\centering\arraybackslash}m{2.2cm}|}
\hline
\textbf{Topic} & \textbf{3GPP group and related documents} & \textbf{Release focus} & \textbf{Section(s)} \\
\hline
\hline
Service requirements and use cases
& SA1, TR~22.837 \cite{tr22837}, TS~22.137 \cite{ts22137}
& Release~19 and Release~20
& Section~II \\
\hline
Channel modeling and target physics
& RAN1, RP~234069 \cite{rp234069}, TR~38.901 sensing extensions \cite{tr38901}, TR~38.765 \cite{tr38765}
& Release~19 and Release~20
& Section~III \\
\hline
Waveform, reference signals, beamforming, and interference
& RAN1, TR~38.765 \cite{tr38765}, and Release~20 contributions
& Release~20
& Section~IV \\
\hline
Protocol signaling, architecture, and exposure
& RAN2, RAN3, TR~23.700-14 \cite{tr2370014}, TS~23.137 \cite{ts23137}, TS~23.138 \cite{ts23138}, ETSI GR ISC~003 \cite{etsiisc003}
& Release~20 studies and Release~21 and later 6G releases
& Section~V \\
\hline
Sensing assisted communication
& Emerging 6G study direction informed by TR~22.870 \cite{tr22870} and TR~38.914 \cite{tr38914}
& Release~20 studies and Release~21 and later 6G releases
& Section~VI \\
\hline
Implementation challenges and open gaps
& Cross group synthesis across SA, RAN, and selected literature
& Release~20 studies and Release~21 and later 6G releases
& Sections~VII and VIII \\
\hline
\end{tabular}
}
\vspace{-0.5cm}
\end{table*}

Existing survey literature on ISAC has extensively addressed waveform design, spectrum sharing, and machine learning techniques \cite{wei2023signals, niu2025interference, luong2026learning}. However, the 3GPP standardization trajectory spanning multiple working groups has not yet been systematically consolidated into a unified narrative. While tutorials exist for specific technologies such as multiple-input multiple-output (MIMO) orthogonal frequency division multiplexing (OFDM) ISAC~\cite{dai2026mimoofdm}, they typically focus on physical layer aspects and do not connect design choices to higher-layer protocol procedures, architecture evolution, and sensing service exposure within a single end-to-end perspective.
This paper addresses that gap by linking service requirements, channel modeling, radio design, and system architecture within one standards-oriented narrative. It distinguishes established specifications from ongoing studies and possible forward directions. It also covers both sensing for environmental perception and sensing-assisted communication, thereby providing a coherent view of the 3GPP trajectory across Service and System Aspects Working Group~1 (SA1), Radio Access Network Working Group~1 (RAN1), Radio Access Network Working Group~2 (RAN2), and Radio Access Network Working Group~3 (RAN3).

This survey paper makes the following key contributions:
\begin{itemize}
\item A standards-centric chronology of ISAC across SA1, RAN1, RAN2, and RAN3.
\item A synthesis of the sensing channel model evolution beyond the legacy TR~38.901 communication baseline.
\item A critical review of waveform design, reference signals, beamforming, and interference tradeoffs in RAN1.
\item A system view of protocol, architecture, and sensing service exposure implications in RAN2 and RAN3.
\item A consolidated discussion of open research challenges and standardization gaps that shape the broader 3GPP evolution toward 6G.
\end{itemize}

\vspace{-0.35cm}
\subsection{Paper Organization}

The remainder of this paper is organized as follows. Section~II examines the standardization landscape, service requirements, KPIs, and sensing topologies. Section~III studies the sensing channel model evolution beyond the legacy TR~38.901 communication baseline. Section~IV reviews RAN1 physical layer design issues, including waveforms, reference signals, beamforming, and interference management. Section~V discusses RAN2 and RAN3 protocol, architecture, and exposure implications. Section~VI examines sensing-assisted communication opportunities. Section~VII synthesizes implementation challenges and open standardization gaps. Section~VIII discusses broader future directions toward 6G. Section~IX concludes the paper.


\vspace{-0.35cm}
\section{Standardization Landscape and Requirements in NR}

Within 3GPP, ISAC is evolving from communication-centric NR toward a managed capability in which sensing can be configured, coordinated, and exposed through the cellular system. Release~19 establishes the foundation through the SA1 feasibility study TR~22.837 \cite{tr22837}, stage~1 requirements in TS~22.137 \cite{ts22137}, and the RAN channel-modeling study item RP~234069 \cite{rp234069}. Release~20 extends this baseline through NR sensing studies and 6G scenario work, including TR~38.765 \cite{tr38765}, TR~22.870 \cite{tr22870}, and TR~38.914 \cite{tr38914}. The transition to normative 6G radio specifications is commonly associated with Release~21 and later releases \cite{romero2025steps}, but feature allocation depends on feasibility, evaluation results, architectural agreement, and standardization priorities. Consequently, ISAC should be viewed as a cross-layer standardization problem: service requirements must be translated into channel models, radio procedures, hardware assumptions, network functions, deployment constraints, and reproducible evaluation methods \cite{baeza2026design}.

Table~\ref{tab:intro_scope_map} maps each survey topic to its 3GPP document family, release context, and section. The discussion follows a simple evidence hierarchy: RP documents indicate approved RAN-level study or work decisions; TRs report study findings, assumptions, alternatives, and evaluation methods; and TSs specify service, architectural, protocol, or radio requirements according to stage and scope. SA1, RAN1, RAN2, and RAN3 respectively cover service requirements, physical-layer aspects, radio-interface procedures, and RAN architecture or interfaces.

\begin{figure*}[!t]
\centering
\includegraphics[width=0.75\linewidth]{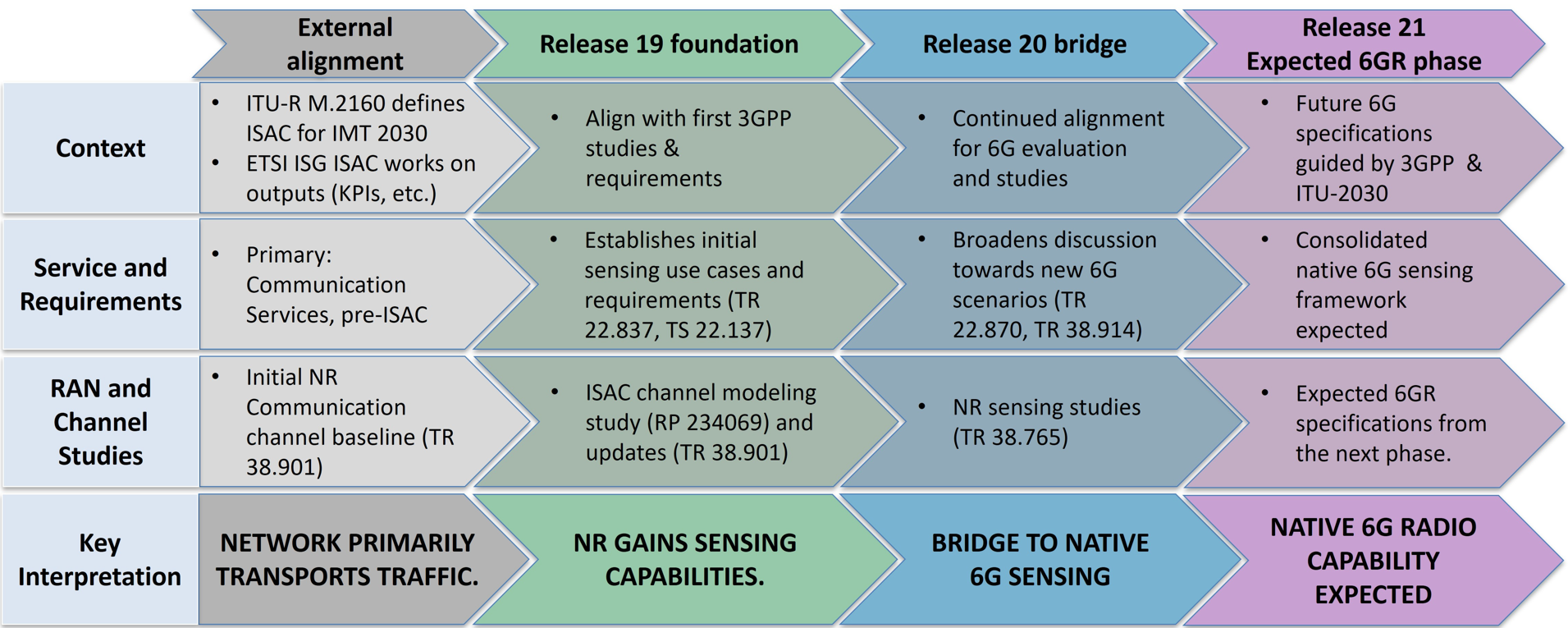}
\caption{Timeline view of the 3GPP ISAC roadmap from NR sensing foundations through Release~20 studies toward Release~21 normative 6G radio work.}
\label{fig:isac_timeline}
\vspace{-0.5cm}
\end{figure*}

Figure~\ref{fig:isac_timeline} summarizes the roadmap from NR sensing foundations to Release~20 studies and expected 6G radio work. ITU and ETSI provide external guidance on use cases, KPIs, channel assumptions, and architecture \cite{iturm2160,etsiisgisac}; within 3GPP, the main anchors are the TR~38.901 sensing-channel extensions \cite{tr38901}, the service foundation in TR~22.837 and TS~22.137 \cite{tr22837,ts22137}, the NR sensing evaluation framework in TR~38.765 \cite{tr38765}, and the broader 6G scenario context of TR~38.914 \cite{tr38914}.


\subsection{3GPP Service Requirements (SA1)}

SA1 defines ISAC as a service in which the cellular system derives information about objects and the environment from radio signals affected by propagation and scattering. TR~22.837 \cite{tr22837} identifies 32 use cases across healthcare, smart cities, transportation, industrial automation, public safety, and related domains, spanning object detection and tracking, environmental monitoring, and motion monitoring. Examples include smart-home intrusion detection, road-safety support, rainfall and flood observation, factory automation, health monitoring, and UAV tracking.

SA1 also defines the service concepts needed for commercial deployment. TR~22.837 distinguishes 3GPP sensing data from processed sensing results, introduces sensing assistance and contextual information, and defines sensing-transmitter (STx) and sensing-receiver (SRx) roles for network nodes or UE \cite{tr22837}. TS~22.137 adds requirements for authorization, operator control, charging, privacy protection, and controlled exposure of sensing results \cite{ts22137}. These requirements show that ISAC cannot be assessed only by detection accuracy, resolution, or range; it must also support managed task establishment, configuration, execution, processing, and delivery. ITU-R~M.2160 and ETSI ISG ISAC provide complementary IMT-2030 use-case, KPI, channel, architecture, and evaluation context, while TR~22.870 extends the discussion to broader 6G services \cite{iturm2160,etsiisgisac,tr22870}.

\subsection{Key Performance Indicators (KPIs)}

ISAC performance combines communication and sensing objectives. Communication metrics such as throughput, spectral efficiency, latency, reliability, coverage, and energy consumption must be preserved because sensing shares the cellular air interface. Sensing metrics include detection probability, missed-detection and false-alarm probabilities, confidence, range/position/angle/velocity accuracy and resolution, sensing latency, refresh rate, coverage, availability, and practical sensing range. Table~\ref{tab:section2_kpis} summarizes these categories and distinguishes 3GPP-grounded metrics from broader research trends \cite{tr22837,ts22137,tr38765}.

\begin{table*}[!t]
\caption{KPI Categories for 3GPP ISAC and Emerging 6G Research Trends}
\label{tab:section2_kpis}
\centering
{
\setlength{\tabcolsep}{3.5pt}
\renewcommand{\arraystretch}{1.15}
\begin{tabular}{|>{\centering\arraybackslash}m{1.15cm}|
                >{\centering\arraybackslash}m{1.2cm}|
                >{\centering\arraybackslash}m{3.8cm}|
                >{\raggedright\arraybackslash}m{5.2cm}|
                >{\raggedright\arraybackslash}m{4.5cm}|}
\hline
\multicolumn{2}{|c|}{\textbf{Category}} & \textbf{Example metrics} & \textbf{Current 3GPP grounding} & \textbf{Emerging 6G research trends} \\
\hline
\hline

\multicolumn{2}{|c|}{Communication} 
& \shortstack[c]{Throughput\\Latency\\Reliability} 
& These remain the primary NR performance objectives. Current Stage~1 sensing specifications do not define a unified set of sensing-specific communication targets across all services \cite{ts22137}. 
& 6G visions aim to sustain ultra-high data rates, low latency, and high reliability while enabling efficient resource sharing between communication and sensing functions \cite{ghosh2025unified,chen2026nextg}. \\
\hline

& Detection 
& \shortstack[c]{Missed detection probability\\False alarm probability\\Detection confidence} 
& TR~22.837 defines missed detection and false alarm as key KPIs. Draft TR~38.765 adopts reference values (e.g., 5\%) for missed detection and false alarm probabilities under specific study assumptions such as UAV scenarios \cite{tr22837,tr38765}. 
& Safety-critical applications motivate stricter detection guarantees, adaptive thresholding, and robustness under dense clutter and interference conditions \cite{ghosh2025unified}. \\
\cline{2-5}

Sensing 
& Estimation 
& \shortstack[c]{Positioning accuracy\\Velocity accuracy\\Range resolution\\Angular resolution} 
& TR~22.837 identifies positioning, velocity, and resolution as key KPI categories. Draft TR~38.765 evaluates baseline targets such as 10~m horizontal accuracy, 10~m vertical accuracy, and 5~m/s velocity accuracy at 90\% confidence for UAV scenarios \cite{tr22837,tr38765}. 
& High-frequency and indoor ISAC studies explore sub-meter positioning accuracy and finer resolution enabled by large bandwidths and advanced signal processing techniques \cite{liu2022isac,baduge2026fr3,dai2026mimoofdm}. \\
\cline{2-5}

& Service 
& \shortstack[c]{Sensing latency\\Refresh rate\\Sensing range\\Coverage continuity} 
& TR~22.837 includes sensing latency and refresh rate as service-level KPIs. However, current RAN1 studies do not yet define universally applicable targets across all deployment scenarios \cite{tr22837,tr38765}. 
& Application-driven studies (e.g., public safety, industrial automation, vehicular sensing) emphasize lower latency, faster update rates, and extended effective sensing coverage tailored to specific use cases \cite{ghosh2025unified,tr22870}. \\
\hline

\end{tabular}
}
\vspace{-0.3cm}
\end{table*}

The NR sensing evaluation framework uses a common KPI family, including detection probability, false-alarm probability, localization accuracy, and velocity accuracy, across target classes and deployment conditions such as outdoor UAV, vehicle, human, indoor AGV, highway, high-speed-railway, and air-to-ground scenarios \cite{tr38765,tr38914,r1_2912,r1_2926,r1_2968,r1_3078}. Service requirements in TR~22.837 and TS~22.137 \cite{tr22837,ts22137} and radio-evaluation assumptions in TR~38.765 and TR~38.914 \cite{tr38765,tr38914} serve different purposes, so scenario-specific KPI values should not be generalized across all use cases. UAV tracking may require long range, fast updates, and accurate 3D position/velocity estimation; indoor presence detection may emphasize detection and false-alarm behavior; and vital-sign sensing requires short-range micro-Doppler sensitivity.

Achievable KPI values depend on waveform structure, reference-signal allocation, antenna aperture, observation duration, propagation geometry, and signal processing. For a CP-OFDM sensing configuration with comb spacing $K_{\mathrm{comb}}$, subcarrier spacing $\Delta f$, $M$ uniformly spaced observations, and interval $T_s$, the monostatic radial unambiguous range is
\begin{equation}
d_{\max} = \frac{c}{2K_{\mathrm{comb}}\Delta f},
\end{equation}
where $c$ denotes the speed of light; increasing comb spacing reduces unambiguous range even if occupied bandwidth still supports fine range resolution. The corresponding maximum unambiguous monostatic radial velocity is
\begin{equation}
v_{\max}=\frac{c}{4f_cT_s},
\end{equation}
where $f_c$ is the carrier frequency, and the Doppler-limited velocity resolution is
\begin{equation}
\Delta v = \frac{c}{2f_cMT_s}.
\end{equation}
These expressions assume uniformly spaced, phase-coherent monostatic observations: smaller $T_s$ increases the unambiguous velocity interval, whereas longer coherent observation time $MT_s$ improves velocity resolution. In bistatic or multistatic sensing, delay and Doppler periodicities are still determined by waveform sampling, but conversion to range and velocity depends on geometry; measured delay corresponds to the transmitter-target-receiver path length, and Doppler depends on target-velocity projection relative to both links. Range resolution is primarily set by effective sensing bandwidth,
\begin{equation}
\Delta r
\approx
\frac{c}{2B},
\end{equation}
where $B$ is the effective contiguous bandwidth. Wider FR3 and higher-frequency bandwidths can enable sub-meter resolution in suitable indoor or short-range deployments \cite{liu2022isac,ghosh2025unified}, but carrier frequency matters mainly through spectrum availability, propagation, antenna dimensions, and hardware capability. Velocity resolution improves with longer coherent Doppler processing, but target acceleration, nonstationary motion, oscillator instability, phase discontinuity, and refresh-rate constraints can cause peak spreading or model mismatch. Position accuracy depends jointly on range, angle, and, where applicable, Doppler information; angular discrimination scales approximately as
\begin{equation}
\Delta\theta
\propto
\frac{\lambda}{D},
\end{equation}
where $\lambda$ is wavelength and $D$ is effective aperture \cite{liu2022isac,ghosh2025unified}. Final position accuracy also depends on SNR, array geometry, calibration, multipath, target extent, estimator design, and multi-node fusion. The KPI framework should therefore distinguish resolution, accuracy, ambiguity, coverage, and service performance: resolution separates nearby targets or parameter values, accuracy measures error to ground truth, ambiguity limits alias-free interpretation, and detection range depends on power, antenna gain, target RCS, propagation loss, clutter, self-interference, and receiver sensitivity. Ali \textit{et al.} \cite{ali2026kpis} provide a broader synthesis of emerging 6G KPIs.

\begin{figure*}[!t]
\centering
\includegraphics[width=0.75\linewidth]{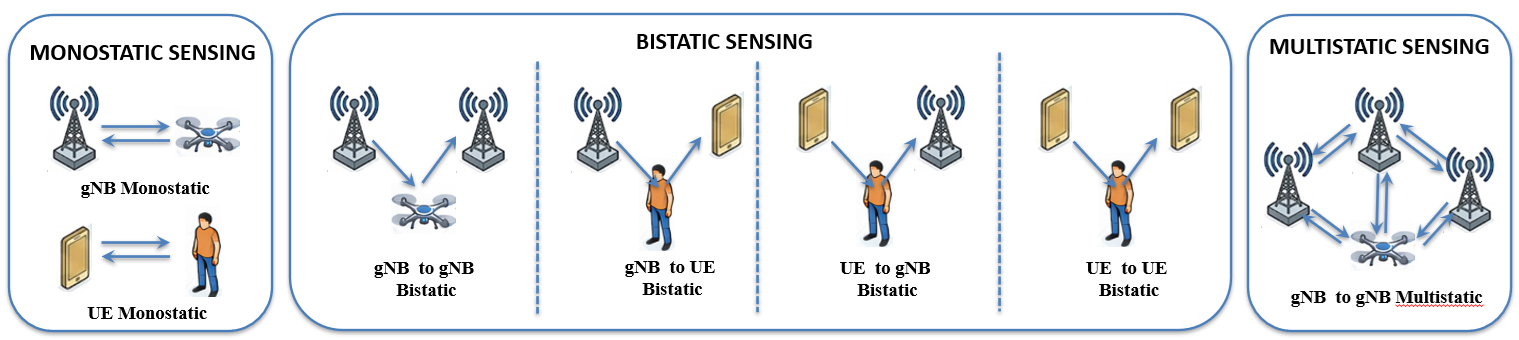}
\caption{Sensing topology considered in 3GPP.}
\label{fig:topology}
\vspace{-0.5cm}
\end{figure*}

\subsection{Sensing Topologies}

Sensing topology follows from the 3GPP STx/SRx roles. Monostatic sensing colocates transmission and reception; bistatic sensing separates them across two nodes; and multistatic sensing coordinates several nodes whose observations may be fused as a sensing group \cite{tr22837,tr38765}. The role-based framework supports gNB monostatic, UE monostatic, gNB-to-gNB, gNB-to-UE, UE-to-gNB, and UE-to-UE sensing, as illustrated in Figure~\ref{fig:topology} and discussed in TR~22.837, TR~38.765, and RP~234069 \cite{tr22837,tr38765,rp234069}. gNB-based monostatic sensing is a practical 5G-Advanced baseline, especially for UAV monitoring, but additional modes enable cooperative perception and public-safety use cases.

Topology also shapes KPI tradeoffs. Monostatic sensing simplifies timing coordination, phase coherence, Doppler processing, and resource reuse, but faces strong self-interference and receiver-isolation constraints \cite{tr38765,ghosh2025unified}. Bistatic sensing reduces self-interference and improves geometric flexibility, but requires accurate time/phase synchronization, geometry calibration, and distributed measurement association \cite{brunner2025bistatic,giroto2026practical,tr38765}. Multistatic sensing adds spatial diversity, blockage robustness, clutter resilience, and wider coverage through fusion \cite{tr22837,ghosh2025unified}, at the cost of greater coordination, transport, signaling, and fusion complexity.

\section{The 3GPP Sensing Channel Model (TR 38.901 Extensions)}
\subsection{Beyond TR 38.901}

The transition from communication-only operation to ISAC changes the role of the propagation channel in a fundamental way. In a communication-centric NR system, the channel is primarily treated as a filter between a transmitter and a receiver, where the objective is to characterize delay spread, angular spread, blockage, and path loss in order to predict link quality. In a sensing-centric system, the channel itself becomes the source of information. Reflected, scattered, and leaked components reveal the presence, location, velocity, and orientation of objects in the environment. For this reason, the Release~19 ISAC study initiated by RP~234069 \cite{rp234069} does not replace the established geometry-based stochastic channel model in TR~38.901 \cite{tr38901}. Instead, it extends this model by introducing a physical object representation together with a target channel, a background channel, and a sensing-specific calibration framework \cite{hong2026channel,yang2026rel19}.

The resulting evolution can be understood as an extension of the legacy geometry-based stochastic model toward a sensing-oriented framework. In the communication-only formulation of TR~38.901, multipath components are generated statistically from scenario-dependent large-scale and small-scale parameters \cite{tr38901}. These parameters describe the link between a transmitter and a receiver. In the sensing extension, the same framework is reorganized into two explicit components, namely a background channel and a target channel \cite{tr38901,hong2026channel,yang2026rel19}. The aggregate ISAC channel can be expressed as
\begin{equation}
H_{\mathrm{ISAC}}(t,\tau)=H_{\mathrm{bg}}(t,\tau)+H_{\mathrm{tg}}(t,\tau)
\end{equation}
where $H_{\mathrm{bg}}(t,\tau)$ represents the static and quasi static environment response, including clutter from buildings, terrain, and surrounding objects, while maintaining compatibility with communication oriented TR~38.901 formulations. In contrast, $H_{\mathrm{tg}}(t,\tau)$ captures the paths that interact directly with the sensing target and therefore carry the target specific delay, angle, Doppler, and reflectivity information required for estimation and tracking \cite{tr38901,hong2026channel}.

\begin{figure*}[!t]
\centering
\includegraphics[width=0.75\linewidth]{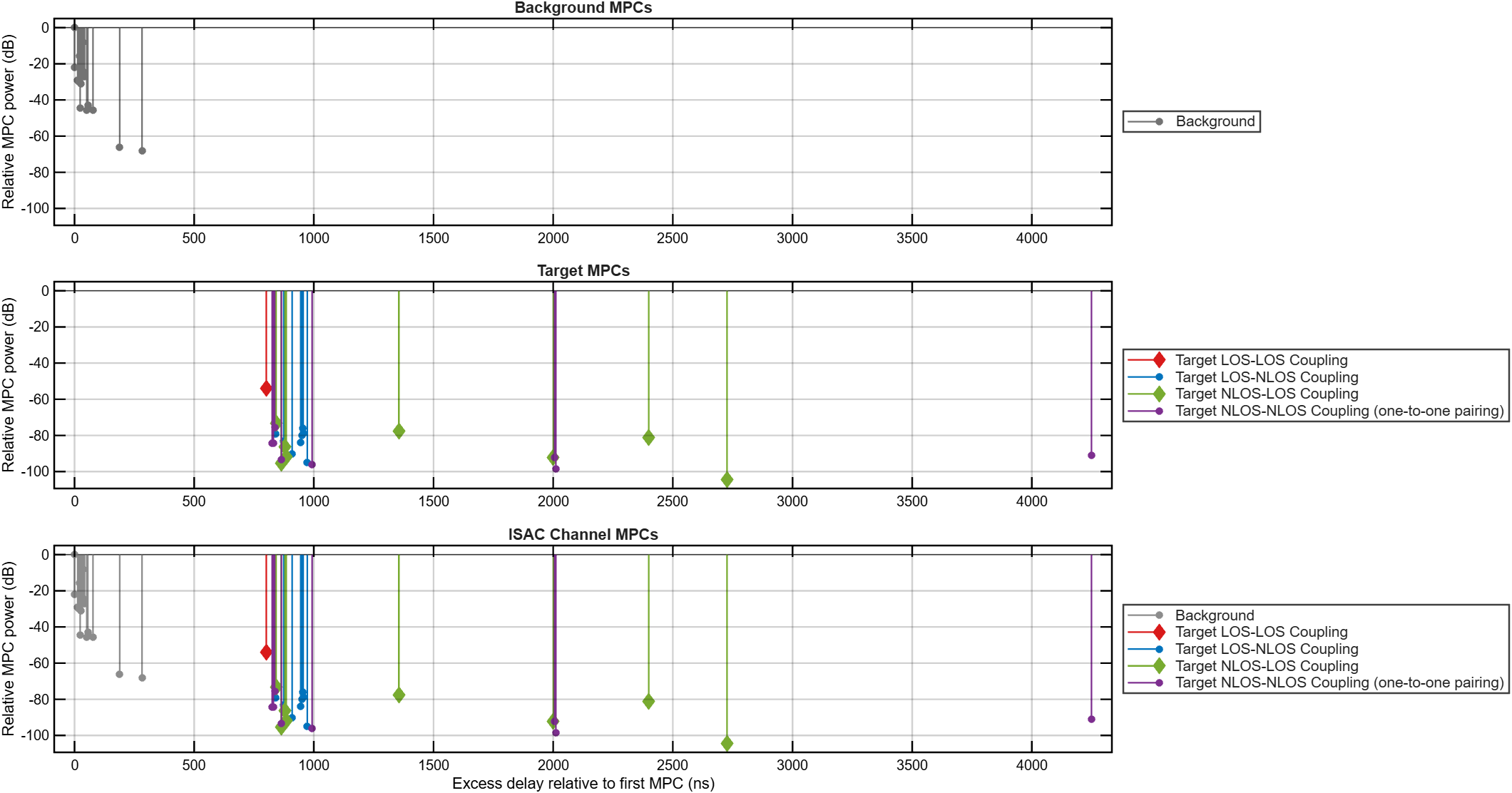}
\caption{ISAC Channel Modeling for UAV in UMa-AV scenario where channel impulse response consists of background and target rays.}
\label{fig:cir}
\vspace{-0.5cm}
\end{figure*}

It is important to mention that the target channel is not generated as a conventional one-hop link. In the TR~38.901 sensing extension \cite{tr38901,hong2026channel,yang2026rel19}, each target-induced component can be interpreted as the concatenation of a transmitter to target link and a target to receiver link, with the target or scattering point acting as the geometric coupling point. As a result, the effective delay, departure and arrival angles, phase, and Doppler are determined by the combined two-leg geometry rather than by a single communication ray. Because of two-leg geometry, the concatenation of transmitter to target link and target to receiver link considers combinations based on their line-of-sight (LOS) and non-line-of-sight (NLOS) propagation states. The two links may appear as LOS to LOS, LOS to NLOS, NLOS to LOS, or NLOS to NLOS combinations, as shown in Fig. \ref{fig:cir}. When both links are NLOS, the synthesized target channel depends on how rays from the two links are associated, which may follow random pairing or a one-to-one pairing rule \cite{tr38901,hong2026channel,yang2026rel19}. For extended targets with multiple scattering points, the target channel becomes a superposition of several such concatenated components. This is why target orientation, point placement, aspect-dependent RCS, and path dropping after concatenation are important for realistic sensing evaluation.

This change also appears in the link budget analysis. Classical communication links follow a one way propagation view in which received power decays approximately with $1/d^2$, consistent with Frii's transmission  equation, where $d$ is the distance between the two communicating nodes. In contrast, sensing links must account for both the illumination path, the scattering from the target, and the collection path at the SRx. For a monostatic or symmetric two-way interpretation, the received echo power follows the radar-equation scaling
\[
P_r \propto \frac{G_t G_r \lambda^2 \sigma}{(4\pi)^3 d^4},
\]
where $\sigma$ denotes the target RCS. In bistatic sensing, however, the range dependence can be described in terms of the sensing-transmitter-to-target distance $d_{\mathrm{STx},\mathrm{tg}}$ and the target-to-sensing-receiver distance $d_{\mathrm{tg},\mathrm{SRx}}$, where STx and SRx denote the sensing-transmitter and sensing-receiver, respectively, giving an approximate dependence of $1/(d_{\mathrm{STx},\mathrm{tg}}^2 d_{\mathrm{tg},\mathrm{SRx}}^2)$ rather than a single $1/d^4$ term. The resulting two-leg loss makes sensing range more sensitive than communication range and explains why bandwidth, antenna aperture, clutter suppression, and residual leakage control become central design dimensions in ISAC studies \cite{ghosh2025unified,hong2026channel,fernandez2026limits}.

This shift from the communication baseline of TR~38.901 to the sensing-oriented extensions used in 3GPP ISAC studies can be described as a continuous transition. In the pre-ISAC baseline, the channel is treated as a propagation filter between transmitter and receiver, path loss follows a one-way decay interpretation, and scatterers are represented through statistical delay and angular spreads. Phase stability mainly supports synchronization and demodulation, while bandwidth and array aperture are treated as communication resources for throughput and beamforming gain. The model focuses on a single link, and unwanted components are handled primarily as interference to be mitigated. In the sensing extensions, the same channel becomes an observation mechanism for targets and the surrounding environment. The link budget must account for two-way echo propagation and explicit target RCS dependence. Targets are described through class, velocity, orientation, and in some cases multiple scattering points. Phase coherence becomes essential for Doppler extraction and motion estimation, while bandwidth and aperture jointly determine sensing resolution in range and angle. The model expands from a single link to monostatic, bistatic, and multistatic configurations. At the same time, structured clutter, direct path leakage, and residual self-interference are treated as explicit channel components because they directly affect sensing performance.

Existing literature on the 3GPP-aligned ISAC channel model spans several complementary directions. The work in \cite{liu2022isac,wei2023signals,lu2024challenges}  explain why legacy communication models are insufficient when target reflectivity, structured clutter, and sensing accuracy are considered. Standards-focused studies in \cite{hong2026channel,yang2026rel19} then examine how Release~19 introduced ISAC-related extensions into TR~38.901 and how these extensions are calibrated and interpreted for practical evaluation \cite{hong2026channel,yang2026rel19}. Beyond this baseline, Heggo \textit{et al.} in \cite{heggo2025etsi} discuss ETSI channel modeling perspectives for RCS and micro-Doppler, Tang \textit{et al.} in \cite{tang2026fr3} provide a tutorial bridge from the Release~19 FR3 specification to simulation practice, G{\'o}mez-Cuba \textit{et al.} in \cite{gomezcuba2026geometry} propose geometry-aware modifications to 3GPP multipath models for bistatic sensing, and Fern{\'a}ndez \textit{et al.} in \cite{fernandez2026limits} analyze the sensing limits of standardized NR waveforms under 3GPP-motivated assumptions. These works indicate a converging view that the 3GPP framework provides a necessary foundation, while richer target physics and geometry-aware sensing assumptions are required for advanced ISAC evaluation.
\subsection{Target Characterization}

The sensing extensions in TR~38.901 \cite{tr38901} introduce a physical object model in which targets are no longer treated as anonymous scatterers. Instead, sensing targets are associated with RCS patterns, orientation states, mobility, and object-dependent geometry. The 3GPP channel model \cite{tr38901} distinguishes target classes such as UAVs, humans, vehicles, and automated guided vehicles, and it defines target channel generation based on a target-specific object model. In practical terms, delay, angle, and power parameters alone are no longer sufficient. They must be coupled with target visibility and aspect-dependent reflectivity \cite{tr38901,hong2026channel}.

TR~38.901 makes this construction explicit at the level of each scattering point of a sensing target (SPST). The RCS of an SPST is defined as a scalar quantity in the local coordinate system of the target and depends on both the incident angle and the scattered angle. The corresponding RCS coefficient for a given pair of angles is expressed as $\sigma_{\mathrm{RCS}}=\sigma_M \sigma_D \sigma_S$, where $\sigma_M$ represents the deterministic component included in the large-scale parameters, while $\sigma_D$ and $\sigma_S$ are included in the small-scale parameters \cite{tr38901}. In this formulation, $\sigma_D$ may either be fixed to 1 or be angle-dependent. For an angle-independent case, such as RCS model 1 with a single SPST, $\sigma_D$ is set to 1. For directional target behavior, the angle dependence is captured through the pattern used in RCS model 2. The third term, $\sigma_S$, follows a lognormal distribution and represents stochastic variation around the deterministic mean behavior. This decomposition separates average reflectivity, directional dependence, and random fluctuation in a manner that remains consistent with the broader TR~38.901 channel generation framework \cite{tr38901,hong2026channel,yang2026rel19}.

As mentioned earlier, the ISAC channel model introduces two RCS models. RCS model 1 represents a small sensing target with a single dominant scattering point and is suitable for compact UAVs and selected human representations. RCS model 2 introduces angle-dependent behavior and can represent either a single larger scattering point or multiple scattering points. For vehicles and automated guided vehicles, the model supports a multiple-point representation in which scattering points are placed at physically meaningful locations such as the front, left, back, right, and roof side of the object. This formulation provides a clear step in which the communication cluster abstraction is reinterpreted in sensing terms as a target with explicit physical dimensions and orientation \cite{tr38901,ghosh2025unified,hong2026channel}.

\begin{figure}[!t]
\centering
\includegraphics[width=1\linewidth]{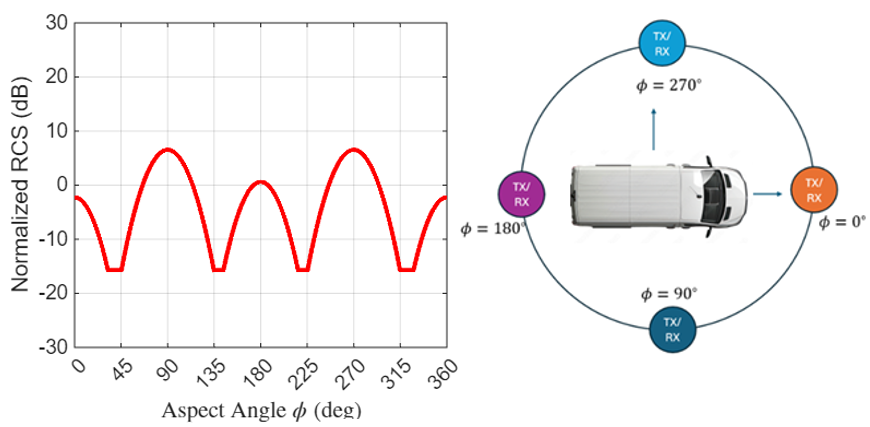}
\vspace{-0.4cm}
\caption{Aspect-angle-dependent normalized RCS pattern for a vehicle target and the corresponding orientation in the TR~38.901 sensing framework.}
\label{fig:vehicle_rcs_pattern}
\vspace{-0.5cm}
\end{figure}

For vehicle targets, this angle-dependent behavior is especially important because the effective RCS varies with the aspect angle of observation. Figure~\ref{fig:vehicle_rcs_pattern} illustrates a normalized RCS pattern as a function of aspect angle together with the associated vehicle orientation. This representation explains why front, side, and rear views contribute differently to the synthesized target channel in the TR~38.901 sensing model \cite{tr38901}.
This also clarifies the distinction between point targets and extended targets. A point target can be approximated by a single dominant scattering center together with an aggregate RCS description. An extended target requires multiple scattering centers because different parts of the object illuminate and reradiate differently across aspect angles and frequencies. The existing literature often relates such RCS fluctuations to classical Swerling type radar models. In contrast, the 3GPP ISAC channel formulation is expressed more directly in terms of target-class-dependent mean behavior and angle-dependent RCS components, rather than adopting a single Swerling label across all scenarios \cite{heggo2025etsi,hong2026channel}. This choice maintains compatibility with the TR~38.901 stochastic framework while allowing future extensions that incorporate richer micro-Doppler effects and material-dependent scattering behavior \cite{chen2019microdoppler,heggo2025etsi}.

\subsection{Environmental Clutter and Spatial Consistency}

The second major extension is the separation between the target channel and the background channel. In a communication setting, the reflected environment is treated as part of the propagation medium and is evaluated through its effect on throughput and reliability. In an ISAC setting, the same environment is interpreted as structured clutter. TR~38.901 therefore constructs the sensing channel as a superposition of target reflections and a background component derived from the communication model. In the bistatic case, the background channel can follow the standard TR~38.901 link generation between a STx and a SRx. In the monostatic case, the model introduces reference points and corresponding virtual link construction so that background backscatter can be represented even when transmission and reception are collocated \cite{tr38901,ghosh2025unified,hong2026channel}.

This separation is essential because clutter is not equivalent to thermal noise. Buildings, ground reflections, vegetation, rails, and other environmental objects may produce strong echoes near zero Doppler and can mask weaker target returns unless suppression methods are applied. The sensing extensions in TR~38.901 therefore include explicit provisions for combining the target channel and the background channel, power normalization across both components, Doppler contributions from mobile scatterers, representation of lower power clusters, and inclusion of additional environmental objects. These provisions maintain compatibility with the legacy NR channel framework while providing a structured basis for evaluating false alarms, missed detections, and clutter limited operation \cite{tr38901,yang2026rel19,ghosh2025unified}.

However, for realistic 6G evaluation, a purely static background assumption is no longer sufficient. In \cite{r1_2915}, it is observed that static clutter can often be mitigated through simple direct-current (DC) removal or zero-Doppler filtering, whereas dynamic clutter created by swaying vegetation, moving non-target objects, or other scene motion raises the residual interference floor and directly affects range-Doppler based target detection. The work in \cite{r1_2915} further notes that UE-based sensing introduces platform-induced non-stationarity, so the clutter field and the useful target returns evolve jointly with the sensing node motion. This perspective is important because it shifts clutter modeling from a largely static background assumption toward a dynamic environment model that is relevant not only for single CPI detection, where the detection decision is made from one coherent observation window without track accumulation across multiple intervals, but also for target tracking across successive observations.

Channel-model development for 6G ISAC extends beyond conventional assumptions of static targets and homogeneous clutter. Several modeling dimensions are important to capture sensing performance more realistically. For instance, target-specific micro-Doppler signatures are relevant for scenarios such as hovering UAVs, human vital signs, gait patterns, and target classification, where bulk Doppler alone may be insufficient for detecting low-velocity or quasi-static objects. In addition, spatial consistency across multiple sensing links, as well as explicit representation of environmental objects (e.g., walls and blockers), play an important role in accurately characterizing propagation and scattering. Environmental effects, including precipitation, fire, and smoke, further introduce time-varying and frequency-dependent distortions that influence sensing reliability. These aspects  collectively indicate that ISAC evaluation should account not only for target range and velocity, but also for micro-motion, cross-link correlation, and dynamic environmental scattering \cite{r1_4280,r1_4710,r1_4889}. Furthermore, analyses across FR3, terahertz, urban-canyon, and multi-band scenarios highlight the impact of dense multipath, background backscatter, and environment-dependent clutter statistics. These factors can significantly influence sensing limits across frequency bands and deployment environments \cite{baduge2026fr3,lyu2026terahertz,chizhik2026backscatter,wang2026multibandfr3}.

Spatial consistency is particularly important in sensing-oriented channel modeling. In communication systems, it ensures that channel parameters evolve smoothly with user displacement, thereby supporting realistic evaluation of beam tracking, mobility, and handover procedures. In sensing systems, spatial and temporal continuity are even more fundamental because successive observations must correspond to the same physical target and surrounding environment. Accordingly, the Release~19 sensing channel framework incorporates spatial consistency into target-channel generation so that target position, orientation, propagation parameters, and clutter characteristics evolve continuously across update instants. Such continuity is necessary for meaningful evaluation of target tracking, Doppler stability, trajectory association, and multi-link fusion; otherwise, abrupt and physically inconsistent channel variations may be incorrectly interpreted as target motion or changes in the environment \cite{tr38901,hong2026channel,yang2026rel19}.

\subsection{High Frequency Effects}

The extension of channel models toward higher-frequency operation introduces additional requirements for ISAC evaluation. TR~38.901 \cite{tr38901} incorporates propagation mechanisms relevant to these bands, including oxygen absorption, wideband channel characteristics, and support for large antenna arrays, together with enhancements for the 7-24~GHz range and sensing-oriented channel generation. These features are particularly important for sensing because increased bandwidth enables finer range resolution, whereas larger array apertures provide improved angular discrimination. At the same time, millimeter-wave and upper-mid-band operation increases sensitivity to blockage, target aspect angle, material properties, and surface roughness, all of which can substantially affect target detectability and parameter-estimation accuracy \cite{tr38901,hong2026channel}.

These propagation characteristics also motivate a more explicit treatment of diffuse scattering. Although dominant interactions may be represented through specular paths in simplified models, rough surfaces, irregular structures, and electrically large objects can produce energy over a wider range of departure and arrival angles. Diffuse and non-specular components therefore influence environmental clutter, target visibility, angular spread, and the stability of sensing returns. Their representation may require material-dependent scattering parameters, surface-roughness descriptions, and segmentation of extended targets into multiple scattering regions, particularly for FR2 and FR3 deployments \cite{heggo2025etsi,ghosh2025unified}. Atmospheric and weather-related losses must likewise be considered according to the operating frequency and deployment conditions. Oxygen absorption is included in TR~38.901, while precipitation-induced attenuation can further reduce the received echo power and practical sensing range at higher frequencies, even when the communication link remains operational \cite{tr38901,ghosh2025unified}.

Beyond the physics-based sensing framework established in Release~19~\cite{tr38901}, AI-assisted and semantic modeling provide complementary approaches for extending channel models to complex and dynamic environments. In this context, machine learning is not intended to replace the physical foundation of the geometry-based stochastic channel model, but rather to support parameter estimation, model calibration, scenario adaptation, and the efficient generation of physically consistent channel realizations. Such data-driven methods can reduce the computational burden associated with exhaustive geometry-specific simulations while preserving relevant propagation constraints and statistical behavior, particularly in dense or highly variable deployment environments \cite{tr22870,tr38914,shatov2025aiml,vaezi2025aiisac,chen2026semantictwin}.

In parallel, semantic channel modeling offers a related abstraction by mapping low-level multipath characteristics to higher-level descriptions of environmental structure, target state, or sensing events. Instead of representing the channel only through path delays, gains, angles, and Doppler shifts, a semantic representation may associate these parameters with objects, activities, blockage conditions, or changes in the surrounding environment. This abstraction can facilitate context-aware sensing, adaptive resource allocation, and intelligent network control without requiring the exchange or processing of complete raw channel-state information \cite{zhang2025channelsemantics,chen2026semantictwin}. These approaches should be regarded as complementary modeling directions rather than replacements for physically grounded channel generation, with their principal value lying in connecting propagation behavior, environmental interpretation, and network optimization.

\section{RAN1: Physical Layer Design for ISAC}

\subsection{Waveform Evolution Toward Joint Design}

At the physical layer, ISAC involves a set of coupled design choices rather than an isolated sensing function. These choices include evaluation assumptions \cite{r1_2915}, communication-sensing integration \cite{r1_2928}, and waveform and frame-structure design \cite{r1_2934,r1_3081}. This joint perspective is essential because communication remains the primary function of the cellular system. For sensing to become a native capability, it must therefore be incorporated into the physical-layer framework without compromising communication coverage, reliability, spectral efficiency, or implementation feasibility. The NR sensing study framework in TR~38.765 \cite{tr38765} provides a basis for evaluating these interactions, including waveform design, reference signals, beam control, and transceiver constraints \cite{r1_2915,r1_2928,r1_2934,r1_3081}. The central design question is consequently not whether sensing can be demonstrated, but which physical-layer operating points can provide useful sensing performance while preserving the robustness and flexibility of the NR communication framework.

Cyclic-prefix orthogonal frequency-division multiplexing (CP-OFDM) provides a natural starting point for cellular ISAC because it is already integrated with NR numerology, scheduling, MIMO processing, synchronization, and radio-frequency implementation. This compatibility enables reuse of existing transceiver architectures, resource grids, and signal-processing procedures with limited modifications. However, CP-OFDM was primarily designed for communication, and compatibility with NR does not necessarily imply optimal sensing performance. Although the OFDM resource grid supports efficient fast-Fourier-transform-based range-Doppler processing, the achievable delay support, Doppler tolerance, dynamic range, and sidelobe suppression depend on the CP duration, subcarrier spacing, windowing strategy, resource allocation, and phase coherence across the CPI.

The ambiguity and sidelobe characteristics of CP-OFDM are particularly important when weak target echoes must be resolved in the presence of strong reflections or environmental clutter. High sidelobe levels may mask weak targets or create spurious detection peaks, whereas irregular allocation of pilots and data resources may further modify the sensing ambiguity function. Echoes with delays beyond the CP duration can introduce inter-symbol interference (ISI), while target-induced Doppler shifts may cause phase variation across OFDM symbols and inter-carrier interference (ICI). These effects can degrade range, velocity, and amplitude estimation, especially for weak or closely spaced targets. Windowing, guard-interval configuration, resource-pattern design, and frame-structure adaptation can mitigate some of these limitations while retaining compatibility with the NR waveform \cite{tr38765,r1_2934,r1_2851,r1_2914,r1_2970,koivunen2025multicarrier,zeng2026ofdm}. Waveform comparisons therefore require evaluation of ambiguity characteristics, coherent-processing assumptions, sensing overhead, and communication degradation under common NR numerology and scheduling constraints, rather than relying solely on isolated radar-oriented metrics \cite{fernandez2026limits,zeng2026ofdm,koivunen2025multicarrier}.

Discrete Fourier transform spread orthogonal frequency-division multiplexing (DFT-s-OFDM) is also relevant, particularly for uplink and UE-transmitted sensing. Its lower peak-to-average power ratio (PAPR) can reduce power-amplifier backoff and energy consumption when a UE acts as the STx. Constant-envelope and constant-modulus waveform designs pursue similar hardware-efficiency objectives and may be useful for power-constrained IoT and UE-based sensing applications \cite{han2026constant}. The suitability of these waveforms, however, depends not only on transmitter efficiency but also on their range-Doppler ambiguity, receiver complexity, resource-grid compatibility, and ability to support communication data and sensing observations simultaneously.

Delay-Doppler-domain waveforms, including orthogonal time frequency space (OTFS), offer another approach for channels with high mobility or strong time-frequency selectivity. Their signal representation is naturally related to propagation delay and Doppler shift, which can facilitate the characterization of target range and velocity in doubly selective environments \cite{nie2026otfs}. On the other hand, affine frequency-division multiplexing (AFDM) and related chirp-based multicarrier techniques provide additional options for high-mobility sensing and operation at higher carrier frequencies. These waveforms are designed to improve resilience to delay-Doppler dispersion while retaining several desirable properties of multicarrier transmission. Their path-separation characteristics may support joint range and Doppler estimation and help distinguish propagation components in rapidly time-varying channels \cite{rou2026afdm,xiao2026afdmrange}. However, such techniques are more appropriately evaluated as alternative waveform candidates rather than treated as inherent components of a baseline cellular ISAC framework. Their applicability depends on synchronization requirements, channel-estimation overhead, spectral containment, transceiver complexity, and coexistence with the communication waveform.

More generally, waveform design for cellular ISAC can be organized around CP-OFDM as a reference configuration, OFDM-compatible enhancements, and alternative waveform candidates. Relevant limitations include monostatic self-interference, PAPR, ambiguity-function sidelobes, spectral leakage, Doppler sensitivity, and limited delay support associated with the CP. Accordingly, waveform assessment requires consideration of the peak-to-sidelobe ratio, integrated sidelobe ratio, PAPR, out-of-band emissions or adjacent-channel leakage, sensing overhead, computational complexity, hardware feasibility, and communication performance. Radar-like or pulsed sensing behavior can also be emulated within an OFDM-compatible framework through appropriate resource allocation, muting strategies, or structured transmission patterns \cite{r1_4326,r1_4712,namgoong2026zipper}.

Data-bearing physical channels may complement reference-signal-based sensing when the SRx has access to the transmitted symbols or can reconstruct them reliably. This approach can increase the usable sensing bandwidth and coherent-processing duration, but it requires clearly defined assumptions regarding symbol knowledge, decoding errors, retransmissions, phase coherence, and communication scheduling \cite{r1_3763,r1_4528}. Similarly, non-uniform sensing-resource allocation and full-duplex waveform design illustrate that delay-Doppler performance, resource-grid compatibility, residual self-interference, and radio-front-end feasibility must be assessed jointly \cite{hong2026otfswaveform,wang2026velocity,wei2026fullduplex}. Therefore, no single waveform metric is sufficient to determine suitability for cellular ISAC; the preferred design must balance sensing accuracy, communication performance, implementation complexity, and compatibility with the overall air-interface architecture.

\subsection{Reference Signal Optimization}

Reference-signal design for ISAC follows an evolutionary path: reuse existing NR signals where possible, enhance them when sensing observability is insufficient, and introduce dedicated sensing signals only when reuse becomes inefficient. Sensing studies often abstract resource allocation through a sensing resource ratio \cite{r1_3079,tr38765}, but operational design must also specify mapping, density, periodicity, beam scheduling, and phase-continuity constraints.

Existing PRS, CSI-RS, SRS, TRS, and DMRS already fit the NR resource grid and configuration framework. PRS is well suited to delay and ranging; CSI-RS and TRS support downlink beam-specific sensing and tracking; SRS provides a natural uplink and UE-assisted basis; and DMRS can aid Doppler estimation when time-frequency density and phase coherence are adequate. Suitability depends on occupied bandwidth, frequency sampling, temporal density, phase continuity, beam coverage, aperture, and calibration, all of which trade against communication resources \cite{tr38765,r1_3079,li2026payload}.

\begin{figure*}[!t]
\centering
\includegraphics[width=0.75\linewidth]{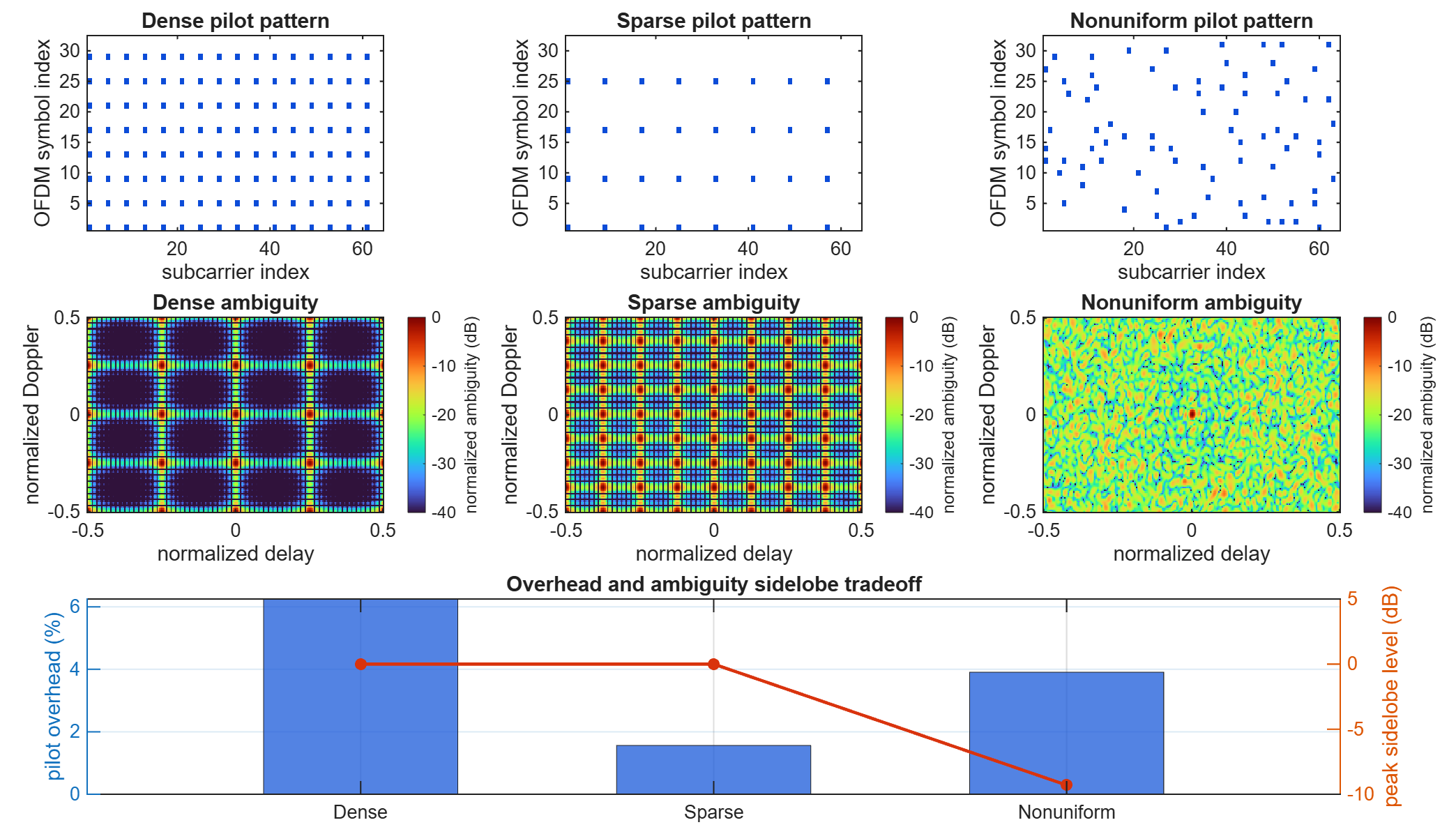}
\vspace{-0.5cm}
\caption{Reference-signal design tradeoff between pilot overhead and two-dimensional ambiguity behavior. Dense pilots improve sensing observability at the cost of higher overhead, sparse pilots preserve resources but introduce stronger ambiguity sidelobes, and nonuniform pilots provide an intermediate design by reducing structured delay-Doppler ambiguities with limited pilot allocation.}
\label{fig:pilotAmbiguity}
\vspace{-0.5cm}
\end{figure*}

The central tradeoff is sensing observability versus communication efficiency. Dense pilots improve delay-Doppler estimation, tracking continuity, and weak-target detection but increase overhead. Sparse periodic pilots preserve resources but can create range/Doppler ambiguities and high sidelobes. Nonuniform pilot placement can reduce structured sidelobes with less overhead than dense allocation, and should be evaluated through the full two-dimensional ambiguity response rather than resource count alone \cite{bouziane2026pilot}. Figure~\ref{fig:pilotAmbiguity} illustrates this tradeoff.

Pilot-aided sensing uses known symbols for matched filtering and coherent integration, making performance less dependent on user traffic but consuming reserved resources. Data-aided sensing reuses payload symbols to enlarge usable bandwidth or observation duration without equivalent pilot overhead \cite{li2026payload}; however, it depends on whether the SRx can access or reconstruct transmitted symbols, scheduling information, retransmission status, and phase coherence. DMRS-assisted payload sensing lies between these cases by using reference symbols for synchronization/calibration and reconstructed data symbols for additional aperture.

Structured NR sequences also support delay estimation, echo separation, and interference discrimination, but communication-oriented configurations may lack sensing bandwidth, density, traffic-independent availability, or multi-slot phase continuity. The resulting design space has three levels: unmodified reuse of existing communication and positioning signals; enhancement through bandwidth, periodicity, comb structure, density, repetition, beam scheduling, or phase-continuity changes; and dedicated sensing reference signals when reuse cannot meet sensing targets without excessive overhead \cite{r1_4307,r1_4754,r1_4711}. PRACH and beam-swept reference signals may also support specialized multistatic localization and angular estimation, provided synchronization, waveform knowledge, beam pointing, calibration, and fusion assumptions are explicit \cite{tosun2026prach,felix2026angular}.

Reference-signal optimization must therefore be tied to sensing procedures. Time-, frequency-, spatial-, code-, and power-domain multiplexing introduce different constraints: clean sensing intervals can interrupt data, simultaneous frequency sharing can reduce contiguous sensing bandwidth, spatial reuse depends on angular separation and array isolation, and code/power multiplexing adds receiver-separation and dynamic-range requirements \cite{r1_5133}. In multi-cell, multi-TRP, or UE-assisted operation, sequence choice, resource coordination, beam scheduling, power control, receiver cancellation, and quiet resources jointly determine observability and interference. Muted or zero-power resources can improve weak-echo reception but consume communication opportunities and may increase latency \cite{r1_3079,r1_2850,r1_2969,huang2026beam}. Table~\ref{tab:section4_nr_rs} summarizes the NR reference signals most relevant to sensing and their limitations.

\begin{table*}[!ht]
\caption{Selected 5G NR Reference Signals Relevant to Sensing.}
\label{tab:section4_nr_rs}
\centering
\setlength{\tabcolsep}{3.6pt}
\renewcommand{\arraystretch}{1.15}
\begin{tabular}{|>{\raggedright\arraybackslash}m{1.4cm}|
                >{\raggedright\arraybackslash}m{3.0cm}|
                >{\raggedright\arraybackslash}m{3.5cm}|
                >{\raggedright\arraybackslash}m{3.9cm}|
                >{\raggedright\arraybackslash}m{4.5cm}|}
\hline
\textbf{Signal} & \textbf{Occupied resources} & \textbf{Frequency-domain pattern} & \textbf{Time-domain pattern} & \textbf{Relevance to sensing studies} \\
\hline
\hline

PRS 
& Configurable bandwidth, typically from 24 to 272 physical resource blocks (PRBs) in steps of 4 
& Comb structures (e.g., 2, 4, 6, 12) 
& Configurable symbol allocation per slot and slot periodicity, with support for consecutive transmissions and muting 
& Strong candidate for delay and ranging estimation due to flexible bandwidth, periodicity, and muting. Primarily downlink-based sensing. \\
\hline

CSI-RS 
& Can span the configured bandwidth part (BWP), including wideband operation 
& Configurable density from one RE per PRB to one RE per two PRBs 
& Typically 1, 2, or 4 adjacent symbols with configurable periodicity 
& Enables beam-aware sensing, angular discrimination, and spatial channel characterization under existing NR configurations. Downlink only. \\
\hline

TRS 
& Spans the configured BWP 
& Comb-4 structure 
& Two symbols with fixed separation and configurable periodicity 
& Supports phase and frequency tracking; however, offers limited flexibility for advanced sensing waveform design. Downlink only. \\
\hline

DMRS 
& Tied to scheduled data resources rather than a dedicated sensing allocation 
& Mapping depends on transmission type and allocation (including comb-based structures) 
& Up to 4 symbols depending on mapping type and scheduling constraints 
& Suitable for opportunistic sensing during data transmission, though performance is constrained by traffic-dependent resource allocation. Supports both uplink and downlink. \\
\hline

SRS 
& Configurable uplink bandwidth from narrowband to wideband PRB allocations 
& Typically 2 or 4 comb structures 
& 1, 2, or 4 symbols with periodic, semi-persistent, or aperiodic transmission 
& Key uplink waveform for UE-assisted sensing and reciprocity-based observation, particularly for uplink probing and environment awareness. Uplink only. \\
\hline

\end{tabular}
\vspace{-0.5cm}
\end{table*}

\subsection{Beamforming and Precoding Strategies}

Beamforming transforms ISAC from passive signal reuse into an actively steerable sensing capability. Conventional communication links generally employ narrow, user-directed beams to maximize received power and spectral efficiency for a selected UE. Sensing, by contrast, may require multi-beam scanning over multiple angular sectors to detect previously unknown objects before assigning dedicated beams for localization and tracking. This difference creates an inherent overhead tradeoff: each beam used for environmental scanning consumes time, energy, and scheduling resources that could otherwise support user data, whereas insufficient scanning reduces visibility outside the directions already served by communication beams \cite{r1_2928,r1_3079,huang2026beam,giordani2019beam}.

This tradeoff motivates joint precoder design, in which the transmit weight matrix is selected to satisfy both communication and sensing objectives. Communication performance is commonly expressed through metrics such as the signal-to-interference-plus-noise ratio (SINR), achievable rate, or user fairness, whereas sensing performance may be characterized through beampattern matching, target illumination gain, estimation error, or the \text{Cramér}-Rao bound. Although fully general real-time optimization may be too complex for direct standardization, these formulations demonstrate that sensing and communication beams cannot always be designed independently \cite{qaisar2026role,dai2026mimoofdm}. Practical beam-management procedures may therefore combine communication-oriented beam refinement with sensing-oriented search, localization, tracking, and adaptation.

Joint beamforming must also account for hardware constraints. Multiuser and multi-target operation may require simultaneous control of transmit gain, inter-user interference, target illumination, and receive sensitivity. Quantized phase shifters, limited radio-frequency chains, nonlinear power amplifiers, phase noise, mutual coupling, and additive transceiver distortions can alter both the achievable sensing SINR and communication quality \cite{he2026beamformingimpairments}. Hardware-aware precoder design is therefore necessary to ensure that theoretical gains remain achievable in practical implementations.

Hierarchical beam search and hybrid analog-digital beamforming provide useful mechanisms for reducing the overhead of exhaustive narrow-beam scanning. A hierarchical procedure can first use broad beams for target discovery and subsequently employ narrower beams for localization and tracking. Hybrid beamforming can further enable multiple simultaneous or adaptively controlled beams while using fewer radio-frequency chains than a fully digital array \cite{giordani2019beam,huang2026beam,dai2026mimoofdm}. The resulting design must balance spatial coverage, array gain, angular resolution, scan latency, energy consumption, and communication throughput.

\subsection{Full-Duplex Operation and Interference Management}

Resource allocation determines how directly sensing and communication functions interact. The two functions may be separated through time- or frequency-domain multiplexing, or integrated by allowing sensing and communication signals to occupy overlapping physical resources \cite{r1_2928,r1_2913,r1_2969}. Separated allocation simplifies receiver processing, interference coordination, and communication quality-of-service management. However, it may interrupt sensing continuity, reduce the target-update rate, and consume dedicated resources. Shared allocation improves resource utilization but introduces additional requirements for receiver processing, power control, scheduling, and link adaptation because the same resources must support communication and sensing objectives with different accuracy and reliability requirements \cite{r1_2850,r1_2913,r1_2969}.

The interference problem is particularly demanding in monostatic sensing, where the STx and SRx are collocated and may operate simultaneously. The receiver must detect weak target echoes in the presence of direct leakage from a substantially stronger transmitted signal. This requires a combination of antenna isolation, propagation-domain suppression, analog cancellation, digital cancellation, and sufficient receiver dynamic range. Residual self-interference may also contain transmitter noise, phase noise, nonlinear distortion, and time-varying coupling components that cannot be represented adequately as stationary additive noise.

Environmental clutter presents an additional structured interference source. Returns from the ground, vegetation, buildings, infrastructure, and other static or slowly varying objects may overlap with the delay, Doppler, or angular characteristics of the target. Consequently, ISAC interference management must jointly address direct-path leakage, residual self-interference, inter-node interference, communication-to-sensing interference, and environmental clutter \cite{tr38765,niu2025interference,wang2026clutter,yan2026fr3dual}. Effective processing may combine calibration, cancellation, clutter-map estimation, background subtraction, adaptive filtering, spatial nulling, and delay-Doppler-angle-domain separation. Full-duplex feasibility therefore depends not only on the total amount of interference suppression but also on where that suppression is achieved. Propagation-domain and analog cancellation are needed to prevent receiver saturation, whereas digital cancellation can remove residual components after analog-to-digital conversion. The required suppression level depends on transmit power, antenna separation, target range, target RCS, clutter strength, waveform bandwidth, and receiver noise figure. These parameters should therefore be specified explicitly when evaluating monostatic ISAC performance.

\subsection{Performance Tradeoff Metrics}

Physical-layer design for ISAC generally defines a Pareto frontier rather than a single universally optimal configuration. Improving one parameter may enhance a particular sensing metric while reducing communication efficiency, increasing hardware complexity, or degrading another sensing dimension \cite{r1_2915,r1_2849,r1_3078}.
A wider effective bandwidth improves range resolution and the ability to distinguish closely spaced propagation paths. However, it also increases sampling-rate requirements, signal-processing load, radio-frequency front-end complexity, and power consumption. Increasing reference-signal density can improve channel observability, Doppler tracking, and velocity estimation, but reduces spectral efficiency by allocating additional resource elements to pilots rather than payload data.

Beamwidth introduces a similar tradeoff. Wide beams provide greater spatial coverage and can reduce the time required for initial target discovery, but they offer lower array gain and poorer angular discrimination. Narrow beams improve target illumination, localization accuracy, interference rejection, and communication link quality, but require more scanning directions and may be sensitive to target motion or beam misalignment.

A longer CPI improves detection sensitivity and Doppler resolution by enabling integration over more observations. At the same time, it occupies resources for a longer duration, increases latency, requires stronger phase coherence, and reduces scheduling flexibility. Target acceleration or trajectory changes may also limit the useful coherent integration time because the constant-delay or constant-Doppler assumptions may no longer remain valid.

Self-interference suppression creates another implementation tradeoff. Greater cancellation improves receiver dynamic range and reduces blind zones in monostatic sensing, but requires more demanding antenna design, calibration, analog circuitry, and digital processing. Similarly, denser sensing-resource allocation can improve estimation accuracy and tracking continuity while increasing communication overhead and inter-node interference.

\subsection{Near-Field and High-Frequency Operation}

Extremely large antenna arrays introduce near-field propagation effects that modify the beamforming assumptions inherited from conventional far-field cellular systems. In the far field, a target is commonly characterized primarily by its direction, and beamforming is based on approximately planar wavefronts. In the radiative near field, spherical wavefronts make the array response dependent on both angle and distance. This property enables joint angular and depth discrimination from a single aperture and supports range-focused or location-focused beamforming \cite{dai2026mimoofdm,chen2026nextg}. Near-field sensing may therefore provide improved localization, spatial separation, and target discrimination, particularly when large apertures are used at short wavelengths. However, it also increases the complexity of channel estimation, codebook design, beam training, calibration, and mobility tracking. Beam focusing must account for both angular and radial dimensions, while the near-field region itself varies with array aperture and wavelength. Models and algorithms based solely on plane-wave assumptions may consequently become inaccurate for sufficiently large arrays or nearby targets.

Operation in upper-FR3 and higher-frequency bands provides larger available bandwidths and electrically larger array apertures, enabling finer delay and angular resolution. These benefits are accompanied by greater propagation loss, stronger blockage sensitivity, increased material and surface dependence, and narrower beams. Atmospheric absorption and hardware limitations may further constrain coverage and receiver sensitivity at very high carrier frequencies \cite{dai2026mimoofdm,tang2026fr3,chen2026nextg,tr38914}.

In these regimes, sensing and communication can support each other directly. Environmental sensing can assist beam discovery, blockage prediction, target-aware beam tracking, and rapid beam recovery, which are described in Section \ref{SAC}. Communication beam measurements can, in turn, provide observations for localization and environmental mapping. The value of this interaction increases as beams become narrower and conventional exhaustive beam sweeping becomes more costly. Consequently, near-field operation, high-frequency propagation, beam management, and sensing-assisted communication should be evaluated jointly rather than as independent physical-layer features.

\section{RAN2 and RAN3: Protocol Stack, Architecture, and System Integration}

While RAN1 addresses the physical transmission and measurement aspects of ISAC, RAN2 and RAN3 are responsible for the protocol procedures, radio-resource coordination, interfaces, and network-level functions needed to configure and coordinate sensing capabilities across the cellular system.

\subsection{RAN2: Protocol Stack and Signaling for ISAC}

The service requirements defined in TS~22.137 \cite{ts22137} and the system-level framework described in TR~23.700-14 \cite{tr2370014} and TS~23.137  \cite{ts23137} establish several capabilities that must be supported by the radio interface, including sensing-task establishment, sensing-role assignment, sensing-input collection, and sensing-result delivery under operator control. From a RAN2 perspective, these capabilities must be translated into signaling procedures that are sufficiently lightweight for dynamic radio operation while remaining expressive enough to configure sensing resources, periodicity, observation duration, accuracy requirements, reporting conditions, and privacy constraints.

An important protocol-design question is whether sensing requires a dedicated connection state or can be supported within the existing Radio Resource Control (RRC) state framework\footnote{The RRC state framework is a control-plane model defined by 3GPP standards (such as 3GPP TS 38.331~\cite{ts38331} for 5G NR) that manages the operational behavior, power consumption, and connection status of a UE.}, including \mbox{RRC\_Connected} and \mbox{RRC\_Inactive}. Reuse of the existing state machine offers the advantage of preserving established mobility, power-saving, security, and radio-bearer procedures \cite{ts22137,tr2370014}. Under this approach, sensing tasks can be introduced through additional RRC configurations and associated lower-layer control signaling. However, the protocol framework must specify which sensing configurations, measurement contexts, and reporting parameters remain valid when a UE changes state, as well as how quickly a suspended sensing task can be resumed.

RRC signaling is suitable for establishing relatively persistent sensing configurations, but sensing operation may also require activation, deactivation, pausing, or adaptation on shorter timescales. This motivates lower-layer control mechanisms that can modify sensing resources, beams, or reporting behavior without requiring complete RRC reconfiguration \cite{jadoon2026architecture,baena2026oran}. ISAC-aware medium access control (MAC) signaling can support such adaptations by controlling sensing opportunities, measurement activation, beam selection, and reporting triggers. Sensing gaps or protected observation intervals may also be configured in a manner analogous to communication measurement gaps, allowing a node to observe echoes or participate in coordinated sensing while selected transmission or reception activities are temporarily modified \cite{tr38765}.

Radio-resource management also needs to balance environmental sensing demands with conventional communication traffic. Persistent sensing tasks, such as UAV surveillance or infrastructure monitoring, may coexist with latency-sensitive and reliability-critical services. The scheduler therefore requires information about sensing-task priority, minimum update rate, observation duration, allowable interruption, and accuracy requirements. This information enables sensing opportunities to be preempted, deferred, shortened, or relocated without violating communication quality-of-service constraints or reducing the sensing refresh rate below its application requirement \cite{tr38765,tr2370014,ts23137}.

This interaction can be represented through a multi-objective scheduling formulation,
\begin{equation}
\max_{\mathbf{s}} \left( w_c R_c + w_s P_d \right)
\quad \mbox{s.t.} \quad
L_c \leq L_{\max}, ;
f_s \geq f_{\min},
\end{equation}
where $\mathbf{s}$ denotes the scheduling decision, $R_c$ represents the communication rate, $P_d$ denotes the sensing probability of detection, and $L_c$ and $f_s$ represent communication latency and sensing refresh rate, respectively. The weighting parameters $w_c$ and $w_s$ express the relative priority assigned to communication and sensing. Predictive or learning-assisted scheduling may provide additional flexibility by allocating sensing resources according to traffic load, target motion, spatial congestion, or expected sensing utility, particularly in high-mobility scenarios such as vehicle-to-everything communication \cite{qaisar2026role,luong2026learning}. Such optimization methods represent implementation options, whereas the protocol framework must primarily expose the measurements, constraints, and control parameters needed to support them.

Sensing measurement reporting naturally follows a layered structure. Compact control-plane reports can convey detection indicators, event notifications, estimated range or Doppler, confidence values, and target-state changes. Larger sensing outputs, including dense delay-Doppler-angle profiles, point clouds, heat maps, or raw in-phase and quadrature samples, may require user-plane transport or dedicated network interfaces. Separating control-oriented information from high-volume sensing data prevents conventional radio-control procedures from being overloaded while preserving timely delivery of latency-sensitive sensing results \cite{tr2370014,ts23137,etsiisc003,jadoon2026architecture}.

TR~38.765 \cite{tr38765} characterizes sensing information through four measurement granularities. Level~A represents low-level channel or waveform measurements similar to channel-state information. Level~B contains processed profiles in the delay, Doppler, and angle domains. Level~C represents detected paths, scattering points, or measurement peaks together with associated attributes. Level~D contains target-level outputs, such as estimated position, velocity, class, or trajectory state. These levels provide a useful framework for relating local sensing processing to reporting overhead and centralized fusion capability.

The four levels can be interpreted as a sensing-information compression hierarchy. Level~A retains the greatest reprocessing flexibility because waveform-level or channel-level information remains available to a centralized sensing function. However, it also imposes the largest requirements in terms of transport bandwidth, latency, storage, energy consumption, and privacy protection. Level~B reduces the payload by transforming measurements into structured parameter domains while retaining surfaces that can support centralized detection and estimation. Levels~C and~D provide progressively more compact reports that are better suited to low-latency control, multi-node coordination, and service exposure, but they require greater local processing and earlier decisions regarding which paths, points, or objects are relevant. The appropriate measurement level therefore depends on the sensing configuration, reporting path, node capability, latency requirement, and desired location of the sensing-processing function. Low-level measurements may be suitable when centralized processing or multi-node coherent fusion provides sufficient benefit to justify their transport cost. Higher-level reports are generally preferable when radio efficiency, UE energy consumption, privacy, and low latency are dominant considerations.

Compact sensing reports must contain sufficient metadata to remain interpretable outside the node that generated them. Relevant information includes measurement timestamps, coordinate systems, uncertainty or confidence values, STx and SRx locations, beam identifiers, synchronization references, and the processing assumptions used to generate the report \cite{tr2370014,ts23137,etsiisc003,jadoon2026architecture}. Without this information, a fusion function may unknowingly combine measurements produced under inconsistent timing, geometry, coordinate-frame, or detection assumptions.

These requirements are particularly important in bistatic and multistatic sensing configurations. When a UE acts as the SRx, forwarding Level~A samples or dense Level~B profiles through the Uu interface\footnote{The Uu interface is the radio interface between a UE and the Next Generation Radio Access Network (NG-RAN), typically a gNB. It carries both user data and control signaling over the wireless link.} can impose substantial signaling and energy costs, even though only a small portion of the measurement space may contain task-relevant information. Levels~C and~D can reduce this burden by allowing the UE or serving node to report selected paths, detections, or target states. However, this approach transfers additional estimation and association responsibility to the reporting node and may reduce the flexibility of centralized fusion \cite{tr38765}.

Bistatic and multistatic reports also require assistance information that is not normally needed for a monostatic measurement. This may include the STx position, SRx position, transmit and receive beam information, timing references, carrier-frequency relationships, and expected delay, Doppler, or angular search regions \cite{r1_4307,r1_4754,r1_4898}. In mobile or unsynchronized deployments, clock offsets, frequency offsets, and node motion can bias delay and Doppler measurements. Consequently, compact reports may remain ambiguous unless synchronization quality, sensing-node geometry, and motion-related metadata are available to the fusion function \cite{bhalli2026doppler}.

RAN2 signaling for ISAC must therefore balance configuration flexibility, reporting overhead, sensing latency, UE power consumption, privacy, and centralized-processing capability. The objective is not to transport the maximum possible amount of sensing information, but to provide the measurement granularity and associated metadata required by the intended sensing task while preserving efficient operation of the radio interface.

\subsection{RAN3: Network Architecture and Interfaces}

Whereas RAN2 addresses radio protocols and signaling between the UE and the radio access network (RAN), RAN3 specifies the architecture, functional distribution, and interfaces required to coordinate sensing across Next Generation Radio Access Network (NG-RAN) nodes. In a communication-oriented network, inter-node interfaces primarily support mobility, load balancing, interference coordination, and bearer continuity. ISAC introduces additional requirements because distributed sensing may involve coordinated transmission, reception, timing, role assignment, and measurement fusion across multiple cells and transmit and receive points (TRPs). Sensing coordination must therefore be treated as an explicit architectural function rather than as a secondary consequence of existing mobility procedures \cite{tr38765,etsiisc003,baena2026oran}.

A central architectural consideration is the placement of sensing control, processing, and analytics functions. ETSI GR ISC~003 \cite{etsiisc003} introduces a sensing control function responsible for handling sensing requests and selecting sensing-data providers, together with a sensing analysis function that processes and fuses collected measurements into sensing results. These entities represent architectural concepts rather than finalized 3GPP network functions, but they illustrate the functional separation required for distributed sensing. Low-latency preprocessing, detection, and local fusion are naturally suited to functions located near the distributed unit (DU) or network edge, whereas multi-site fusion, historical analysis, policy coordination, and service-level orchestration needs to be performed at the central unit (CU), edge cloud, or core network \cite{jadoon2026architecture}.

A disaggregated RAN architecture provides one possible realization of this functional split. Policy definition, long-term optimization, and sensing-service intent can be associated with the Non-Real-Time RAN Intelligent Controller (Non-RT RIC), whereas near-real-time task adaptation, beam control, resource coordination, and sensing-aware radio optimization can be supported through Near-Real-Time RAN Intelligent Controller (Near-RT RIC) applications. Sensing analysis can be distributed across local processing functions, the Near-RT RIC, edge-computing resources, and the Non-RT RIC according to latency, data-volume, and fusion requirements. Under such an interpretation, interfaces such as A1 and E2\footnote{The A1 interface connects the Non-RT RIC and Near-RT RIC, supporting policy guidance, enrichment-information exchange, and AI/ML-related coordination. The E2 interface connects the Near-RT RIC to the underlying RAN nodes and enables near-real-time measurement collection, monitoring, and control of RAN functions and resources.} may convey sensing policies, measurement-quality indicators, timing and geometry information, and control actions for sensing-aware radio operation \cite{etsiisc003,baena2026oran,jadoon2026architecture}. 

The Xn and F1 interfaces\footnote{The Xn interface interconnects NG-RAN nodes and supports inter-node mobility, coordination, signaling, and user-plane forwarding. In contrast, the F1 interface connects the centralized and distributed components of a disaggregated gNB, with F1-C supporting control-plane signaling between the gNB-CU-CP and gNB-DU and F1-U supporting user-plane transport between the gNB-CU-UP and gNB-DU.} similarly require sensing-aware information exchange. Over Xn, neighboring gNBs may need to coordinate sensing roles, timing references, resource allocations, beam configurations, interference constraints, and measurement reporting. Over F1, the CU and DU may exchange sensing-task configurations, activation status, processing instructions, and intermediate sensing information between centralized control and distributed execution. These capabilities are particularly relevant to bistatic and multistatic operation, in which one node transmits while another node receives. In such configurations, reliable fusion depends on coordination among the participating nodes in terms of scheduling, waveform and resource use, timing, beam direction, and measurement association \cite{etsiisc003,baena2026oran,jadoon2026architecture}.

Inter-cell resource coordination also acquires an additional role in ISAC. Communication-oriented coordination primarily addresses throughput, handover, and inter-cell interference, whereas sensing transmissions introduce strong probing signals or broad beams that reduce the sensitivity of neighboring SRxs. Coordinated scheduling, beam planning, transmit-power control, sensing-resource partitioning, and interference management therefore become important architectural functions \cite{tr38765,etsiisc003,baena2026oran}. These mechanisms need to address both communication-to-sensing interference and interference among simultaneous sensing tasks.

\begin{figure}[!t]
\centering
\includegraphics[width=0.5\linewidth]{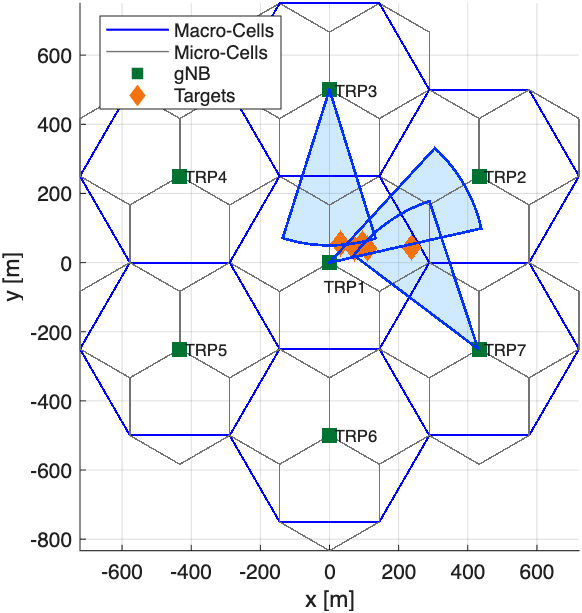}
\vspace{-0.4cm}
\caption{Multi-TRP cooperation in which multiple gNBs coordinate their sensing activities in an urban macro aerial-vehicle (UMa-AV) scenario.}
\label{fig:multiTRP}
\vspace{-0.5cm}
\end{figure}

Multi-TRP cooperation is particularly valuable in NLOS, cluttered, or high-accuracy scenarios. Observations from spatially separated nodes can provide geometric diversity, reduce blockage sensitivity, improve angular coverage, and support more robust localization and tracking \cite{tr38765}. However, the associated performance gains depend on synchronization, coordinate alignment, measurement uncertainty, target association, and the consistency of local processing assumptions. Figure~\ref{fig:multiTRP} illustrates a multi-TRP configuration in which several gNBs coordinate their sensing activities in an urban macro aerial-vehicle scenario. In \cite{varshney2026multitrp}, we demonstrate that single-TRP sensing can suffer from high missed-detection rates because of limited motion observability. Multi-TRP-assisted sensing substantially improves detection performance and localization robustness, while also enabling reconstruction of the target’s true three-dimensional velocity vector.

\subsection{System-Level Integration and Core Network Interaction}

System-level ISAC integration extends beyond the RAN. The service requirements, architectural studies, and system specifications described in TR~23.700-14 \cite{tr2370014}, TS~23.137 \cite{ts23137}, and TS~23.138 \cite{ts23138} provide a framework in which sensing can be requested, authorized, configured, processed, and exposed to network and external consumers. Under this framework, sensing is not only an internal radio capability but also a network service whose execution may involve radio access, edge processing, core-network control, and application-level consumption.

Sensing as a service can therefore be interpreted as an exposed network capability rather than as a fixed RAN procedure. ETSI GR ISC~003 considers both internal network consumers and external third-party consumers of sensing information \cite{etsiisc003}. The service-based architecture of the 5G system provides a possible foundation for controlled exposure. For example, an external sensing request may be received through a network exposure mechanism, authorized according to operator policy, and translated into a sensing task with corresponding geographic, temporal, accuracy, and resource constraints. The Network Exposure Function (NEF) specified in TS~23.501 \cite{ts23501} provides a possible exposure point for such requests. Under this arrangement, service exposure is handled separately from the radio-resource execution required to realize the sensing task \cite{ts23138}.

The service perspective also clarifies the separation between control-plane and data-plane functions. The control plane is suited to sensing-task authorization, service-area configuration, performance-target definition, sensing-role selection, privacy enforcement, and charging control, as reflected in TS~22.137 \cite{ts22137}. High-volume sensing information is more appropriately transported through the user plane or through a dedicated sensing-data path. Such data includes delay-Doppler-angle profiles, point clouds, heat maps, target tracks, or raw in-phase and quadrature samples for centralized or edge processing \cite{etsiisc003,jadoon2026architecture,baena2026oran}. Separating task control from sensing-data transport prevents large measurement payloads from overloading control signaling and allows the transport mechanism to be selected according to latency, reliability, and bandwidth requirements.

Privacy, security, and trust are fundamental to sensing-service exposure. Sensing results can reveal object location, human activity, movement patterns, environmental conditions, or other sensitive information that is not normally contained in conventional communication traffic. Sensing services therefore require authorization, authentication, access control, data minimization, retention policies, charging, and role-based exposure \cite{ts22137,tr22837,etsiisc003}. The sensing system also need to restrict measurement resolution, geographic coverage, or result granularity according to the identity and privileges of the requesting consumer.

The progression from communication-oriented protocols to an ISAC-aware system can be summarized across several layers. At the RRC layer, connection management, mobility control, and measurement configuration can be extended with sensing-task parameters, role assignment, authorization, reporting periodicity, accuracy requirements, and persistence of sensing context across RRC states. At the MAC layer, scheduling must support rapid sensing activation, sensing gaps, beam coordination, resource multiplexing, and dynamic adjustment of the sensing duty cycle in response to communication demand.

Across Xn and F1, procedures for mobility, load balancing, and distributed control can be supplemented with timing alignment, sensing-role coordination, beam and resource information, and transfer of sensing-related measurements and metadata. At the core-network and service-exposure layers, sensing requests must be authorized, translated into executable radio tasks, and mapped to participating sensing entities. At the data-plane level, conventional communication flows may be complemented by sensing products such as detected paths, target states, point clouds, range-Doppler profiles, and, where justified, lower-level measurement samples for downstream fusion and analytics.

End-to-end sensing operation therefore requires coordination among service exposure, core-network control, RAN configuration, radio scheduling, measurement processing, and sensing-result delivery. The architecture must specify not only where sensing functions reside, but also how sensing tasks are decomposed, how participating nodes are selected, how measurements are associated with a common coordinate and timing framework, and how results are delivered to authorized consumers.

In summary, RAN2 and RAN3 provide the protocol and architectural mechanisms required to transform ISAC from a physical-layer measurement capability into an integrated network function. Radio-resource management, signaling procedures, inter-node coordination, functional placement, and service exposure ultimately determine whether sensing and communication can coexist efficiently and whether distributed observations can be converted into reliable and usable sensing services.

\section{Sensing Assisted Communication}
\label{SAC}
Once sensing is available as a managed network capability, its utility extends beyond target detection and environmental perception. Spatial, temporal, and propagation-related observations can be incorporated into communication procedures to reduce control overhead, improve link robustness, and support anticipatory adaptation. This reciprocal relationship is particularly relevant to FR2, FR3, and higher-frequency operation, where narrow beams, rapid channel variation, and abrupt blockage can reduce the effectiveness of communication procedures that rely primarily on reactive measurements \cite{tr38914,tr22870,qaisar2026role,chen2026nextg}.

Sensing-assisted communication encompasses several physical-layer and radio-resource-management functions, including beam management, channel state information (CSI) acquisition and reporting, channel estimation, mobility and handover support, link adaptation, power saving, and transmit-power control. For these functions, the required sensing output is not necessarily a final object label or a complete environmental map. In many cases, a compact representation of the dominant propagation structure is sufficient, including multipath delays, Doppler shifts, angles, powers, blocker locations, reflecting surfaces, or the estimated position and motion of relevant objects. This perspective motivates procedure-level evaluation metrics such as beam-selection success probability, beam-alignment latency, reference-signal overhead reduction, channel-estimation error, and beamforming gain, rather than relying exclusively on end-to-end throughput metrics that may depend strongly on scheduler implementation and traffic assumptions \cite{r1_4307,r1_4816,r1_4710,r1_4889}.

\subsection{Sensing-Aided Beam Management}

\begin{figure}[!t]
\centering
\includegraphics[width=0.75\linewidth]{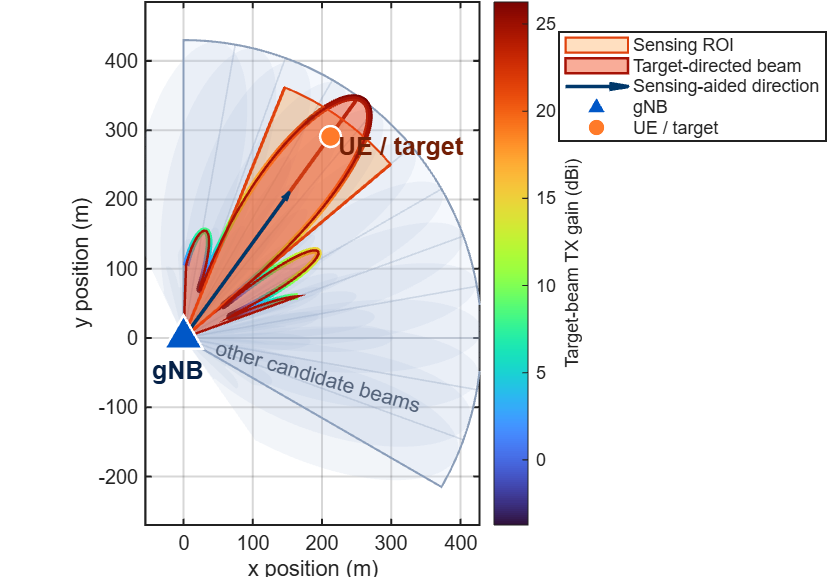}
\vspace{-0.3cm}
\caption{Illustration of sensing-aided beam management, where sensing information narrows the beam search to a region of interest and reduces exhaustive sweeping overhead.}
\label{fig:beamROI}
\vspace{-0.5cm}
\end{figure}

Beam management is one of the most direct beneficiaries of sensing support. In highly directional systems, conventional beam training often relies on exhaustive or near exhaustive sweeping across many candidate beams, which consumes significant time and frequency resources and becomes costly in dense mobility scenarios. If the gNB can infer the position, direction, or motion trend of a user from sensing observations, beam discovery can be limited to a much smaller candidate set, as illustrated in Figure~\ref{fig:beamROI}. 

In this way, sensing transforms beam management from a largely blind search process into a geometry-guided procedure and reduces the control overhead associated with synchronization signals, channel state information reference signals, and beam refinement stages \cite{giordani2019beam,huang2026beam,qaisar2026role}.

The same principle extends to beam tracking. By observing the velocity and trajectory of a vehicle, drone, or handheld user, the network can predict the direction of the dominant beam in the next interval and update the beam or precoder before significant misalignment occurs. This predictive beamforming approach is particularly effective in scenarios with fast motion, body shadowing, or street-level blockage, where beam coherence time is short. In such conditions, sensing provides motion priors that complement feedback-driven beam refinement and improve tracking reliability \cite{huang2026beam,tr38914,chen2026nextg}.

A key prerequisite for these sensing-assisted loops is that the sensing result remains fresh over the prediction window used by the communication controller. If $\tau_s$ denotes the sensing observation time, $\tau_p$ the processing and reporting delay, and $\tau_a$ the actuation delay for beam or resource reconfiguration, then the predicted user or blocker state must remain valid at time $\tau_s+\tau_p+\tau_a$. In practice, this requires the sensing refresh period and end-to-end control latency to be smaller than the beam coherence time or blockage evolution time scale. Otherwise, stale sensing information can shrink the beam search space around an outdated direction and degrade beam tracking rather than improving it. Therefore, sensing-assisted communication should be evaluated not only by beam-selection success or pilot-overhead reduction, but also by prediction-window validity, sensing-result age, motion-model uncertainty, and the robustness of the assisted procedure to delayed or imperfect sensing inputs \cite{huang2026beam,qaisar2026role,chen2026nextg}.

\subsection{Channel Estimation and Prediction}

Channel estimation and prediction constitute another major benefit of sensing support. Traditional CSI acquisition relies on repeated pilot transmission to observe a channel that may vary rapidly in time, frequency, and angle. Under ISAC, part of the channel geometry can instead be inferred from the surrounding environment. A sensing enabled network can construct a radio environmental map (REM) that identifies dominant reflectors, likely blockage sources, and user trajectories, and then combine this information with sparse pilot observations to predict the evolution of dominant multipath components \cite{jadoon2026architecture,qaisar2026role,li2026payload}.

This geometry awareness remains valuable even when the direct path is blocked. If sensing identifies a stable reflecting surface, such as a building facade or another persistent scatterer, the network can intentionally exploit a NLOS path rather than discovering it only after repeated pilot failures. In this way, sensing does not replace CSI acquisition, but it makes channel prediction more physically grounded and reduces the density of reference signaling required to maintain reliable operation in large array and wideband systems \cite{tr38914,chen2026nextg,luong2026learning}.

\subsection{Proactive Handover and Blockage Prediction}

\begin{figure*}[!t]
\centering
\includegraphics[width=1\linewidth]{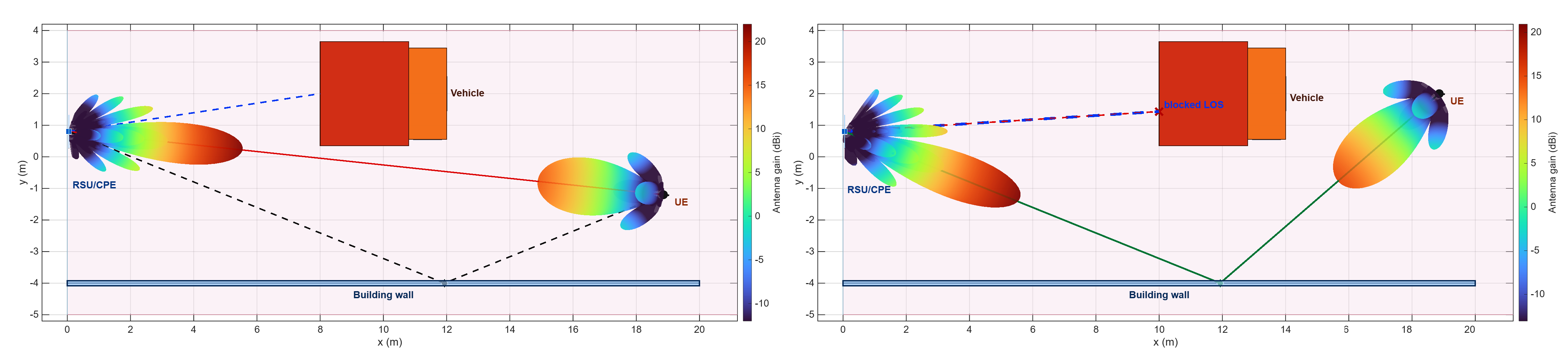}
\vspace{-0.8cm}
\caption{Illustration of sensing-assisted proactive mobility support. The sensing function detects an approaching vehicle that blocks the direct RSU/CPE--UE link before link failure occurs, enabling the network to redirect the beam through an alternative reflected path from a building wall.}
\label{fig:blockage}
\vspace{-0.5cm}
\end{figure*}

Proactive mobility support is another key opportunity enabled by sensing. At high carrier frequencies, links can degrade abruptly when pedestrians, buses, trucks, or turning vehicles intersect the dominant propagation path. A sensing-aware network can detect such approaching blockers before the received signal quality collapses, and can therefore adjust beam directions, activate backup links, or initiate handover earlier than a purely reactive mobility procedure would allow \cite{giordani2019beam,qaisar2026role,tr22870}. Figure~\ref{fig:blockage} illustrates this principle for a vehicle-blockage case. In the initial state, the RSU/CPE maintains a direct beamformed link to the UE while sensing identifies the vehicle intersecting the dominant propagation path. Once the direct path becomes vulnerable, the network can proactively select an alternative reflected path from the building wall, thereby maintaining connectivity without waiting for a reactive beam failure event.

This concept extends beyond individual blockers to broader traffic and crowd dynamics. If the network observes increasing congestion in an intersection, corridor, or venue, it can preconfigure mobility parameters, reserve alternative resources, or redistribute traffic across layers and cells before service degradation becomes visible. This anticipatory capability distinguishes sensing-assisted communication from conventional communication approaches and highlights its importance in the evolution toward 6G systems \cite{tr22870,tr38914,chen2026nextg}.

\subsection{Sensing-Aided Resource Allocation and MAC Efficiency}

Sensing also improves resource allocation at the MAC layer. Spatial awareness can guide the selection of beams, pilot density, modulation and coding schemes, and scheduling periodicity in a way that reflects the actual motion and blockage conditions of the environment. When user motion is predictable, the network can reduce probing overhead and allocate more resources to data transmission. When environmental uncertainty increases, the scheduler can provision denser reference signaling or activate diversity mechanisms before packet losses occur \cite{li2026payload,luong2026learning,qaisar2026role}.

These benefits are particularly important for sidelink and vehicular operation, where hidden nodes and rapidly changing geometry complicate decentralized coordination. If vehicles or infrastructure nodes can sense nearby objects and predict future resource occupancy, packet collisions can be reduced even when some contenders have not yet transmitted. Sensing priors can also shorten initial access by narrowing the angular search space for synchronization and random access, thereby reducing access delay in dense or highly directional deployments \cite{tr22870,chen2026nextg,qaisar2026role}. Table~\ref{tab:section6_compare} summarizes the broader transition from sensing-blind communication procedures to sensing-assisted communication.

Sensing also enhances interference management. When the network can detect the position, motion, or angular characteristics of dominant interferers and strong reflectors, it can adjust beam directions, apply spatial nulling, control transmit power, and refine scheduling decisions before interference significantly affects reception. This capability is especially valuable in dense deployments, where interference patterns depend on mobility, blockage, and dynamic scene geometry. In such scenarios, sensing-assisted interference control complements conventional channel-state-information-driven coordination by introducing explicit environmental awareness, which can improve effective signal to interference plus noise ratio and stabilize link quality under challenging propagation conditions \cite{niu2025interference,qaisar2026role,chen2026nextg}.

\begin{table*}[!t]
\caption{Evolution from Sensing Blind Communication to Sensing-Assisted Communication}
\label{tab:section6_compare}
\centering
{
\setlength{\tabcolsep}{3.8pt}
\renewcommand{\arraystretch}{1.15}
\begin{tabular}{|>{\raggedright\arraybackslash}m{2.7cm}|
                >{\raggedright\arraybackslash}m{5.2cm}|
                >{\raggedright\arraybackslash}m{6.5cm}|}
\hline
\textbf{Feature} & \textbf{Legacy 5G communication practice} & \textbf{Sensing-Assisted 6G direction} \\
\hline
\hline

Beam training 
& Exhaustive or codebook-based beam sweeping with significant control overhead in highly directional operation 
& Directed or predictive beam search guided by sensed user position, motion, and environmental geometry (e.g., reflectors and blockers) \\
\hline

Handover and beam switching 
& Largely reactive adaptation triggered after signal degradation or beam failure 
& Proactive adaptation enabled by detection of blockers, trajectory prediction, and anticipated link degradation \\
\hline

CSI acquisition 
& Pilot-intensive estimation with frequent probing to track channel variations 
& Reduced pilot overhead through refinement aided by environmental priors, sensing feedback, and REMs \\
\hline

Link establishment under blockage 
& Recovery initiated after dominant path loss, often incurring additional delay and signaling overhead 
& Alternative or reflected paths identified in advance using sensed geometry, enabling faster recovery and reduced disruption \\
\hline

MAC resource allocation 
& Scheduling driven primarily by traffic demand and channel feedback with limited environmental awareness 
& Scheduling incorporates motion dynamics, blockage prediction, and scene awareness to adapt pilots, beams, and diversity mechanisms \\
\hline

Initial access 
& Broad spatial search for synchronization and random access in directional systems 
& Sensing-assisted narrowing of the search space, reducing access latency and overhead \\
\hline

Link reliability 
& High-frequency links are sensitive to blockage and beam misalignment 
& Reliability improved via prediction, early reconfiguration, and sensing-informed spatial control \\
\hline

\end{tabular}
}
\vspace{-0.5cm}
\end{table*}
Overall, these mechanisms show that sensing-assisted communication is an important direction for 6G systems because environmental awareness can reduce control overhead, enable earlier adaptation, and strengthen resilience against blockage and interference. 

\section{Implementation Challenges and 3GPP Constraints}

\subsection{Hardware Impairments and Architectural Constraints}

The radio front end is a fundamental constraint for monostatic ISAC because a gNB must illuminate the environment and detect weak echoes while transmitting through a colocated or closely coupled radio chain. Echoes may be many orders of magnitude weaker than direct leakage, antenna coupling, and transmitter distortion, so practical sensing depends on antenna isolation, propagation-domain suppression, analog/digital cancellation, receiver dynamic range, linearity, and calibration stability \cite{tr38765,fernandez2026limits,yan2026fr3dual}.

Self-interference suppression is distributed across the transceiver. Antenna separation, polarization isolation, circulators, and beam-domain nulling reduce coupling before reception; analog cancellation prevents receiver saturation; and digital cancellation removes residual components after sampling. The usable sensing dynamic range depends on the combined suppression and on variations with beam direction, transmit power, temperature, hardware aging, and nearby objects, which require adaptive calibration.

Power-amplifier and RF impairments further limit sensing fidelity. Higher transmit power can extend range, but amplifier saturation introduces nonlinear distortion, spectral regrowth, and in-band intermodulation that may violate emission limits or obscure weak echoes. Phase noise, frequency offset, sampling-clock error, IQ imbalance, and array gain/phase mismatch broaden delay-Doppler peaks, bias range/velocity estimates, distort beampatterns, and degrade angular accuracy \cite{tang2026fr3,ghosh2025unified,fernandez2026limits}. Unlike communication receivers, which can often tolerate residual distortion through coding and equalization, sensing relies on coherent phase relationships across subcarriers, symbols, antennas, and observation intervals. Hardware calibration and impairment compensation must therefore be part of the sensing model.

Architecture determines which impairments dominate. Fully digital arrays support flexible cancellation and calibration but require many RF chains and high-resolution converters. Hybrid arrays reduce complexity and power while limiting simultaneous communication beamforming, sensing illumination, and self-interference suppression. Separate transmit/receive panels improve isolation but increase size, calibration complexity, and cost; time-separated sensing avoids simultaneous coupling but reduces sensing continuity and resource efficiency.

\subsection{Spectrum, Coexistence, and Regulatory Constraints}

Spectrum regulation constrains ISAC because sensing transmissions must coexist with NR traffic, adjacent-band services, incumbent systems, and national or regional emission rules. ITU-R SM.1541 \cite{itursm1541} specifies out-of-band unwanted-emission limits, but compliance becomes harder when sensing requires high power, wide bandwidth, rapid beam switching, or waveforms with unfavorable spectra. Practical deployments may need guard bands, spectral shaping, tighter filtering, amplifier backoff, reduced duty cycle, or lower transmit power, trading sensing range, update rate, or resolution against regulatory compliance \cite{tang2026fr3,chen2026nextg}.

Coexistence must be evaluated at network scale. A probing signal acceptable within one cell may desensitize neighboring receivers, disrupt communication traffic, or affect incumbents; broad beams, high-power illumination, and coordinated multi-node sensing can enlarge the interference footprint. Conversely, neighboring-cell or incumbent transmissions can raise the sensing noise floor and mask weak echoes. Coordination across time, frequency, space, power, and waveform domains is therefore needed through sensing-resource partitioning, inter-cell scheduling, beam restrictions, transmit-power control, interference-aware waveforms, or protected observation intervals.

Shared and unlicensed spectrum add channel-access constraints. Listen-before-talk or similar procedures can delay, shorten, or prevent sensing transmissions, interrupt CPIs, reduce periodicity, and complicate distributed coordination \cite{threegpplaaupdate}. Incumbent protection may also require metrics beyond average received interference, including peak emissions, spatial duty cycle, aggregate multi-node interference, antenna orientation, beam occupancy, and the probability of illuminating a protected receiver.

\subsection{Privacy, Security, and Legal Constraints}

Privacy and security are stringent in ISAC because sensed entities may not be authenticated devices or subscribers. TS~22.137 identifies authorization, privacy protection, and controlled exposure as essential requirements, building on TR~22.837 \cite{ts22137,tr22837}. Unlike conventional communication-data processing, device-free sensing may infer the presence, location, movement, behavior, or physical characteristics of entities with no signaling relationship to the network, raising requirements for lawful purpose, consent where applicable, data minimization, geographic and temporal scope, retention, and disclosure control \cite{ts23137,ts23138}.

Protection must cover the full sensing-information lifecycle: task authorization, participating-node selection, measurement collection, local processing, reporting, fusion, storage, and exposure. Access control may depend on purpose, service area, granularity, target category, result accuracy, and consumer identity. The product granularity also matters: raw samples and delay-Doppler-angle representations may enable secondary inference, while object-level reports can reveal identity-related attributes or trajectories. Data minimization must therefore limit content, resolution, duration, and spatial coverage to the authorized task.

ISAC expands the threat surface beyond signaling and user-plane attacks. Waveform manipulation, spoofed or replayed observations, false-target injection, malicious environmental modification, falsified coordinates, and corrupted sensing-assistance data can affect beam selection, handover, resource allocation, autonomous control, or safety services even when communication bearers remain cryptographically protected. TS~33.501 provides authentication, confidentiality, integrity, and secure network-function interaction \cite{ts33501}, but cryptographic protection of the reporting path does not prove the physical validity of a sensing observation.

Sensing reports should therefore be bound to contextual metadata such as the sensing task, allocated resource, observation and reporting times, STx/SRx identities, node locations, beam configuration, coordinate frame, and processing level. Integrity protection of this metadata enables replay detection, temporal-validity checks, geometry-based plausibility testing, and multi-node consistency checks before sensing information is used for beam steering, handover, safety services, infrastructure control, or external exposure \cite{ts23137,ts23138,yang2026security,qu2026privacy}. Distributed consistency is useful but not decisive because synchronization error, blockage, clutter, calibration mismatch, and local algorithms can also produce disagreement.

Governance becomes more complex when sensing products leave the operator network. Vertical applications, infrastructure operators, transportation systems, industrial platforms, public-safety agencies, and other consumers may need different access levels. TS~23.138 and the service-based architecture in TS~23.501 provide foundations for controlled exposure, which should include authorization, purpose limitation, auditing, charging, retention control, restrictions on onward transfer, and data-sovereignty constraints on where raw or processed measurements may be stored or transferred \cite{ts23138,ts23501}.

Privacy risk can also arise at the physical layer: shared communication/sensing waveforms may allow unintended receivers to infer transmitter location, channel structure, or surrounding geometry even when payloads are encrypted. Countermeasures include waveform and beam control, artificial noise, access-controlled reference signals, spatial suppression, randomized resources, and controllable propagation \cite{kumar2026transmitterprivacy}, with tradeoffs in communication quality, sensing accuracy, power, and complexity. Evaluation should consider unauthorized sensing range, inference accuracy, report-integrity failure probability, spoofing/replay resilience, exposure granularity, retention duration, auditability, and attacker capabilities.

\subsection{Standardization Gaps and Deployment Constraints}

Cellular sensing standardization must balance technical capability with implementation complexity. Initial 5G-Advanced studies often prioritize gNB-based sensing because it enables feasibility evaluation without immediately imposing major UE hardware, battery, uplink, or reporting changes. The monostatic UAV baseline in TR~38.765 provides controlled radio assumptions \cite{tr38765}, while TR~23.700-14 and TS~23.137 address coordination of sensing entities, tasks, inputs, and results \cite{tr2370014,ts23137}.

These baselines do not yet cover all sensing configurations. UE-based sensing, bistatic and multistatic operation, multi-TRP cooperation, cross-cell fusion, persistent services, and external exposure require further procedures for synchronization, role assignment, UE capability signaling, power consumption, measurement transport, privacy, and distributed processing. Transport and functional placement are especially important because F1, Xn, and open fronthaul were not designed for continuous raw samples, dense delay-Doppler-angle profiles, point clouds, or other high-rate intermediate sensing products.

The functional split must determine which operations run at the radio unit, distributed unit, central unit, edge, or core. Low-latency tasks such as interference suppression, matched filtering, detection, and local feature extraction may stay near the radio site; multi-node association, trajectory estimation, historical analysis, and service orchestration may be centralized. Intermediate sensing data may need compression into path-, point-, or object-level representations before network transport \cite{etsiisc003,baena2026oran,jadoon2026architecture,tr2370014,ts23138}.

The main standardization gaps therefore cluster around hardware/RF feasibility, protocol and architecture procedures, spectrum/coexistence compliance, and privacy/security governance. These dimensions are coupled: centralization can improve fusion but increases transport, latency, energy, and privacy exposure; local processing reduces volume but depends on node-specific algorithms and calibration; stronger sensing improves range but increases self-interference, emissions, and coexistence risk. Table~\ref{tab:section7_gaps} summarizes the gaps that constrain scalable sensing beyond baseline NR configurations.

\begin{table*}[!t]
\caption{Open Challenges and Standardization Gaps for ISAC}
\label{tab:section7_gaps}
\centering
{
\setlength{\tabcolsep}{3.6pt}
\renewcommand{\arraystretch}{1.15}
\begin{tabular}{|>{\raggedright\arraybackslash}m{2.7cm}|
                >{\raggedright\arraybackslash}m{4.0cm}|
                >{\raggedright\arraybackslash}m{4.4cm}|
                >{\raggedright\arraybackslash}m{4.1cm}|}
\hline
\textbf{Area} & \textbf{Current 3GPP status} & \textbf{Gap} & \textbf{Future direction} \\
\hline
\hline
\textbf{Channel modeling}
& Partial. RP~234069 \cite{rp234069}, TR~38.901 extensions \cite{tr38901}, and draft TR~38.765 \cite{tr38765} provide the present baseline
& Limited treatment of micro-Doppler, extended target physics, clutter benchmarking, and reproducible concatenation assumptions
& Extended target and scene models with explicit benchmark profiles and clearer reporting of author assumptions \\
\hline
\textbf{Waveform and reference signals}
& Baseline in the OFDM family under draft TR~38.765 \cite{tr38765} and ongoing RAN1 studies
& Communication-first designs remain sensing inefficient and can create excessive pilot overhead or reduced sensing fidelity
& Joint waveform and reference signal design with flexible time and frequency mapping \\
\hline
\textbf{Self-interference control}
& Assumed in the monostatic baseline of draft TR~38.765 \cite{tr38765}
& Hardware limits in isolation, cancellation depth, dynamic range, and calibration still constrain practical deployment
& Practical full-duplex-aware design across radio-frequency hardware, baseband processing, and conformance requirements \\
\hline
\textbf{Multi-node coordination}
& Evolving through RAN2 and RAN3 studies together with ETSI GR ISC~003 \cite{etsiisc003}
& Synchronization, fusion timing, data volume, and cross cell coordination remain difficult under NLOS and clutter
& Multi-TRP cooperation, coordination, and sensing fusion support across several nodes \\
\hline
\textbf{RAN signaling and reporting}
& Evolving. TS~22.137 \cite{ts22137}, TR~23.700-14 \cite{tr2370014}, and draft TS~23.137 \cite{ts23137} and TS~23.138 \cite{ts23138} define the current direction
& No unified sensing-aware MAC and RRC procedures yet exist, and scalable layered reporting remains incomplete
& Dedicated sensing-aware procedures with adaptive reporting granularity and clearer service interfaces \\
\hline
\textbf{Service exposure and trust}
& Defined at service and system-level in TS~22.137 \cite{ts22137}, TS~23.501 \cite{ts23501}, TS~23.138 \cite{ts23138}, and TS~33.501 \cite{ts33501}
& Enforcement of consent, minimization, provenance, charging, and lawful exposure remains unclear in practical deployment
& Trusted sensing as a managed service with policy-driven exposure, provenance control, and governance \\
\hline
\textbf{AI-native scene understanding}
& Exploratory in TR~22.870 \cite{tr22870}, TR~38.914 \cite{tr38914}, and existing literature
& No standardized framework exists for semantic sensing outputs, model lifecycle management, closed-loop learning, or validation
& AI-integrated sensing control with semantic scene abstraction and closed-loop learning tied to network decisions \\
\hline
\textbf{Non-terrestrial and aerial sensing}
& Exploratory in TR~22.870 \cite{tr22870}, TR~38.914 \cite{tr38914}, and existing literature
& Limited study of joint terrestrial and aerial sensing workflows, continuity, and coordination across domains
& Unified terrestrial and non-terrestrial sensing support with continuity across domains \\
\hline
\end{tabular}
}
\vspace{-0.5cm}
\end{table*}

\section{Research Directions for 6G ISAC}

The evolution of cellular sensing involves several technological directions that extend beyond the baseline capabilities considered in initial ISAC frameworks. These directions can be interpreted as research and standardization opportunities supporting the evolution toward 6G ISAC. Their common objective is to transform sensing from an isolated radio measurement function into an intelligent, distributed, and resource-aware capability integrated across the radio interface, network architecture, computing infrastructure, and service layer.

\subsection{AI-Native ISAC and Semantic Intelligence}

A central research direction is deeper integration of sensing, communication, computing, and AI across the radio interface and network architecture. IMT-2030, TR~22.870, and TR~38.914 describe networks with greater environmental awareness, distributed intelligence, and context-dependent operation \cite{iturm2160,tr22870,tr38914}. AI can support target classification, clutter suppression, parameter estimation, model calibration, and joint sensing/communication control \cite{qaisar2026role,chen2026nextg,shatov2025aiml,vaezi2025aiisac}. An AI-native architecture would decide which measurements are needed, where they should be collected, which nodes should participate, and how sensing accuracy should be balanced against throughput, latency, energy, and signaling overhead. 

Such control requires measurement value rather than measurement quantity alone. The network may allocate more sensing resources to uncertain geometry, moving objects, or blockage-prone regions, while reducing updates in stable areas. Semantic intelligence provides a complementary abstraction by converting raw samples, channel coefficients, point clouds, or delay-Doppler-angle profiles into task-relevant descriptions such as pedestrian movement, traffic congestion, approaching blockers, occupied safety regions, or infrastructure changes \cite{zhang2025channelsemantics,chen2026semantictwin}. These representations should retain confidence, uncertainty, location, observation time, and validity duration so downstream functions can verify, fuse, or reinterpret them.

Agent-based and generative architectures provide possible coordination mechanisms. Policy-governed agents may reason over communication, sensing, computing, and service requirements across network entities \cite{ferrag2026agents}; generative models may combine sparse observations with learned environmental priors to reconstruct missing measurements or candidate scene states \cite{chen2026semantictwin}. Compact latent or semantic representations can reduce transport of full point clouds or wideband samples \cite{polese2026dapps,zhang2025channelsemantics}, but generated information must be distinguished from directly observed evidence. Evaluation should therefore include physical consistency, uncertainty calibration, inference latency, generalization, computational cost, failure behavior, and the effect of incorrect AI decisions on sensing and communication.

\subsection{Real-Time Digital Twins and Radio Environmental Maps}

Radio environmental maps and digital twins provide a structured mechanism for maintaining knowledge of the physical and propagation environment. A radio environmental map may contain information about coverage, dominant paths, blockers, reflecting surfaces, interference sources, and mobility patterns. A digital twin extends this concept by maintaining a dynamic representation that evolves with changes in objects, infrastructure, traffic, and radio conditions \cite{zhang2026digitaltwin,chen2026semantictwin}. Such a representation enables the network to anticipate communication and sensing events rather than reacting only after conventional measurements indicate degradation. For example, the digital twin may predict when a dominant path will become blocked, which reflected paths are likely to remain available, where sensing coverage is insufficient, or which beams and TRPs are suitable for a particular task \cite{tr22870,tr38914}. This capability supports proactive beam management, handover preparation, resource allocation, sensing-node selection, and interference coordination.

The digital twin can also represent the condition of the network infrastructure itself. Changes in observed beam patterns, path geometry, or calibration parameters may indicate antenna misalignment, mechanical drift, environmental damage, or hardware degradation \cite{zhang2026digitaltwin}. The network therefore becomes capable of observing not only users and external objects, but also the physical condition of its own deployment. Digital twins are especially relevant to industrial automation, transportation, smart infrastructure, and public-safety applications, where both physical processes and radio connectivity evolve over time \cite{tr22870,chen2026nextg}. They can support counterfactual evaluation by estimating the effect of changing a beam, activating another TRP, modifying a sensing schedule, or rerouting a communication link before the action is applied to the physical network.

Goal-oriented optimization provides an additional use of the digital twin. Rather than maximizing sensing fidelity uniformly, the network may update only those parts of the twin that are relevant to the intended task. For example, a beam-selection procedure may require reflector and blocker geometry but not detailed classification of every object. Task-specific updating can reduce sensing, processing, and transport overhead while preserving the information required by the consuming application \cite{saggese2026digitaltwin}.

The value of a digital twin depends on its synchronization with the physical environment. Relevant performance measures therefore include spatial accuracy, update latency, prediction horizon, consistency across sensing nodes, uncertainty, and the rate at which the representation becomes outdated. A detailed but stale twin may be less useful than a coarser representation that is updated reliably and with low latency. Scalable implementation also requires decisions about data ownership, functional placement, and model partitioning. Local twins may be maintained near individual cells or sites, whereas wider-area representations may be constructed at edge or centralized functions. The architecture must determine how local updates are fused, how inconsistent observations are resolved, and which parts of the representation can be exposed to external applications.

\subsection{Reconfigurable Intelligent Surfaces, Near-Field Operation, and Higher-Frequency Sensing}

Reconfigurable intelligent surfaces (RISs) provide a programmable mechanism for modifying the propagation environment. By controlling the phase, amplitude, polarization, or direction of reflected fields, an RIS can strengthen selected communication paths, improve sensing illumination, or extend coverage into regions that are poorly visible from the serving node \cite{wu2025rissurvey}. This capability is particularly relevant when blockage and weak diffraction limit direct propagation. In an ISAC configuration, an RIS may support both communication and sensing. It can redirect probing energy toward a region of interest, enhance echoes from a weakly illuminated target, create a controllable NLOS path, or improve geometric diversity for localization. Multiple surfaces may further provide complementary viewing directions or coverage across complex environments \cite{li2026multiris}.

However, RIS-assisted sensing introduces several modeling and implementation challenges. The network must estimate or calibrate the cascaded transmitter-surface-target-receiver channel, distinguish RIS-controlled paths from uncontrolled multipath, and coordinate surface states with waveform, beam, and resource allocation. Passive surfaces may have limited control and sensing capability, whereas active or semi-passive surfaces introduce additional power consumption, noise, hardware complexity, and regulatory considerations.

Extremely large antenna arrays introduce a related transition from conventional far-field operation to radiative near-field propagation. In the far field, beamforming is commonly described through direction-dependent plane waves. In the near field, wavefront curvature makes the array response dependent on both angle and distance \cite{dai2026mimoofdm,dai2026nearfield}. This property enables range-focused or location-focused beams and permits joint angular and depth discrimination from a single aperture. Near-field ISAC may therefore improve three-dimensional localization, separation of targets with similar angles but different ranges, and spatial control of communication and sensing energy. However, it also increases the complexity of channel modeling, codebook design, beam training, target tracking, and array calibration. A beam must be configured over both angular and radial dimensions, and a location-focused beam may become ineffective when the target moves outside its focal region.

The boundary between near- and far-field behavior depends on wavelength, array aperture, and propagation distance. As carrier frequency increases and array dimensions become electrically larger, near-field effects can arise over distances relevant to cellular operation. Spherical-wave modeling and joint aperture design should therefore be incorporated into evaluations whenever plane-wave assumptions are no longer accurate \cite{tang2026fr3,dai2026nearfield}.

Upper-FR3, sub-terahertz, and terahertz operation can provide large bandwidths and fine delay and angular resolution. These advantages are accompanied by increased propagation loss, stronger blockage sensitivity, molecular absorption, surface-dependent scattering, narrower beams, and more demanding radio-frequency hardware \cite{chen2026nextg,tr38914}. Higher-frequency sensing should consequently be evaluated jointly with beam management, near-field propagation, hardware impairments, atmospheric effects, and realistic material properties.

RISs, extremely large arrays, and higher-frequency operation are therefore closely related. Larger apertures and programmable surfaces can compensate for some coverage limitations of short-wavelength propagation, while sensing information can assist their configuration. At the same time, their benefits depend on calibration, channel knowledge, control latency, and the ability to coordinate a large number of spatial degrees of freedom.

\subsection{Non-Terrestrial Networks  and Ubiquitous Integrated Sensing}

The integration of sensing with non-terrestrial networks (NTNs) extends environmental perception across terrestrial, aerial, and spaceborne platforms. Satellites, high-altitude platform stations, unmanned aerial platforms, and terrestrial nodes can contribute complementary observations with different coverage areas, viewing geometries, revisit times, and propagation conditions. This architecture can support applications such as maritime monitoring, agriculture, disaster assessment, wildfire observation, infrastructure supervision, and wide-area transportation management \cite{iturm2160,tr38914}. Non-terrestrial sensing can also strengthen positioning, navigation, and timing. When conventional satellite-navigation signals are obstructed, degraded, or unavailable, communication and sensing signals from terrestrial and non-terrestrial nodes may provide complementary range, Doppler, angle, synchronization, and environmental information \cite{chen2026nextg,tr22870}. Fusion across these sources can improve resilience, although it requires consistent coordinate systems, timing references, and uncertainty models.

The architecture differs substantially from terrestrial-only ISAC. Satellite motion produces rapidly varying geometry, large Doppler shifts, long and asymmetric propagation delays, and time-varying coverage. Payload power, processing capacity, antenna aperture, feeder-link capacity, and onboard storage also constrain the amount of sensing data that can be collected and transported.

In cooperative multi-satellite sensing, several satellites may jointly serve communication users and observe targets through distributed geometries \cite{kim2026multisatellite}. Centralized fusion can exploit measurements from multiple platforms but requires transport of observations to a common processing function. Distributed processing reduces signaling overhead and latency but may limit access to complete measurements and complicate consistency across satellites.

Bistatic LEO configurations provide another architecture in which the satellite illuminates or communicates while a terrestrial gateway receives the target echoes \cite{zhang2026leobistatic}. Such separation may reduce selected onboard receiver requirements or provide favorable sensing geometry, but it introduces demanding synchronization, beamforming, power-allocation, and calibration requirements. The rapidly changing transmitter-target-receiver geometry must be represented explicitly in both the sensing model and the fusion algorithm.

Gateway placement and satellite visibility also influence sensing observability. A target may be illuminated by a satellite but not visible to the intended terrestrial receiver, or the bistatic geometry may produce poor range or velocity sensitivity. Network planning should therefore consider sensing geometry in addition to conventional communication coverage.

Handover acquires additional meaning in an NTN ISAC system. The system may need to transfer not only a communication session, but also a sensing task, accumulated target state, environmental context, and measurement history across beams, satellites, gateways, or terrestrial nodes \cite{bhandari2026handover}. Maintaining continuity requires common timing, coordinate frames, task identifiers, and uncertainty representations.

Research on satellite and NTN ISAC consequently spans waveform design, Doppler compensation, cooperative sensing, distributed fusion, beam management, handover, payload architecture, and satellite-terrestrial interfaces \cite{jamshed2026ntn,wei2026leo}. Evaluation must account for orbital motion, platform constraints, atmospheric propagation, observation geometry, revisit intervals, and the communication overhead required to coordinate sensing across segments.

\subsection{Sustainable and Resource-Frugal ISAC}
\begin{figure*}[!t]
\centering
\includegraphics[width=0.75\linewidth]{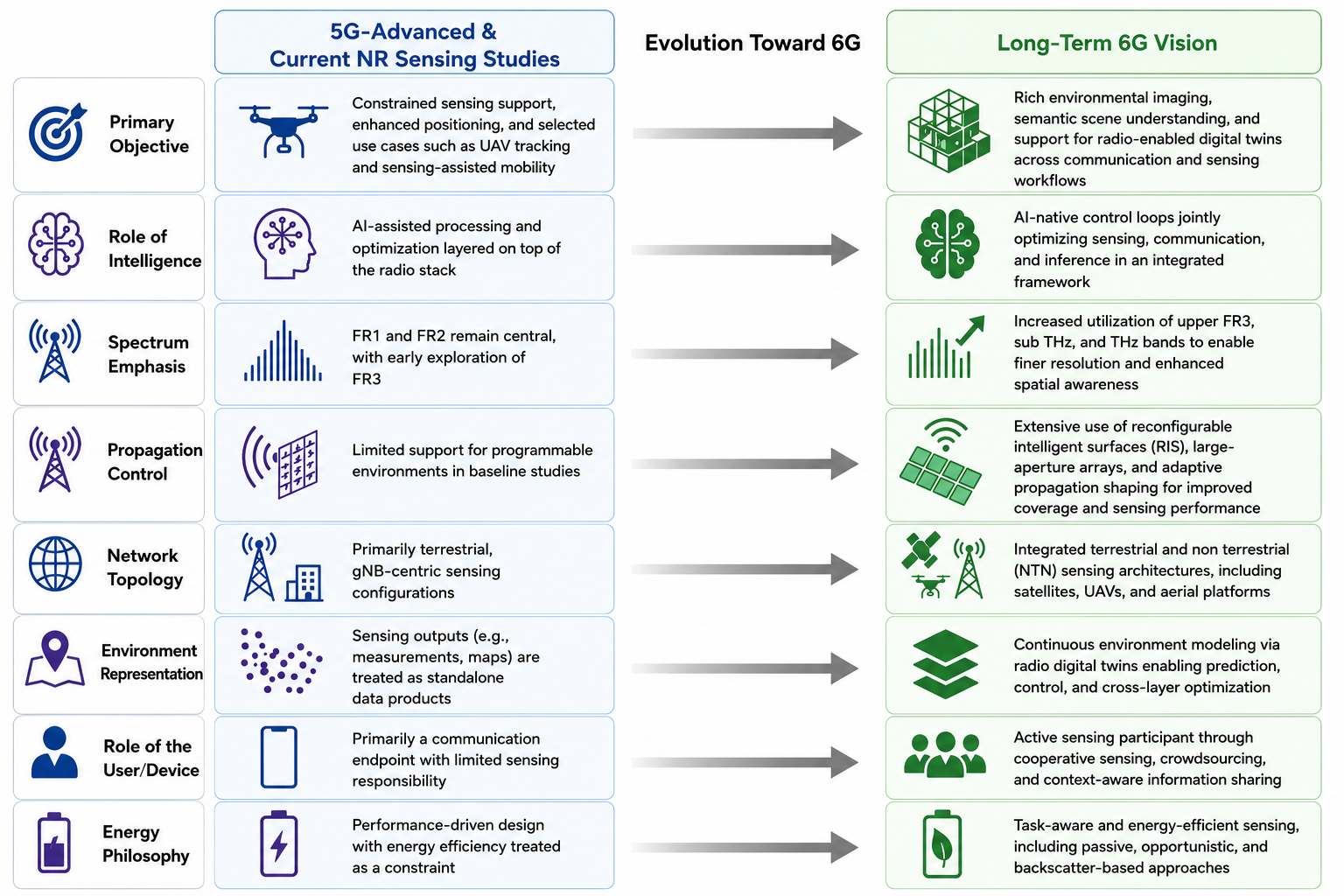}
\vspace{-0.4cm}
\caption{Progression from baseline cellular sensing capabilities toward broader 6G ISAC research directions across intelligence, environmental representation, propagation control, non-terrestrial integration, and sustainability.}
\label{fig:section8_compare}
\vspace{-0.5cm}
\end{figure*} 

Sustainability is a fundamental design dimension for large-scale sensing networks. Not every task requires maximum bandwidth, continuous illumination, dense reporting, or centralized processing. Sensing fidelity should instead be adapted to the application requirement and environmental uncertainty. Presence detection, blockage warning, occupancy estimation, and high-resolution environmental reconstruction, for example, impose substantially different demands on waveform bandwidth, update rate, angular coverage, and processing complexity \cite{chen2026nextg,qaisar2026role,tr22870}. A resource-frugal architecture should therefore allocate the minimum sensing effort required to meet a specified confidence, latency, or accuracy target. The network may adapt sensing bandwidth, coherent processing duration, beamwidth, duty cycle, node participation, measurement granularity, and reporting level. Stable environments may be observed less frequently, whereas regions with rapid motion or uncertain geometry may receive additional resources.

Task-oriented sensing also reduces unnecessary data transport. A blockage-warning service may require only a predicted event, location, confidence value, and validity interval rather than a complete point cloud. Similarly, local processing may convert raw measurements into compact target- or event-level reports before transmission. Such compression reduces fronthaul load and energy consumption, although it can limit centralized reprocessing and should preserve uncertainty and provenance information.

Energy consumption must be considered across the complete sensing chain, including signal transmission, radio-frequency operation, analog-to-digital conversion, local processing, data transport, fusion, storage, and AI inference. A configuration that reduces transmit power but requires continuous high-rate sample transport may not reduce the total system energy. Sustainability evaluation should therefore use end-to-end energy measures rather than transmit power alone.

Low-power and energy-neutral devices provide another research direction. Backscatter-assisted ISAC can enable passive or nearly passive devices to convey information and interact with sensing signals without conventional active transmission \cite{zhang2025netzero}. Such devices may support inventory monitoring, environmental sensing, identification, or low-rate status reporting with limited battery dependence. Backscatter-based sensing nevertheless introduces challenges related to weak signal strength, ambient interference, channel variability, device identification, synchronization, and separation of direct and backscattered components. Its usefulness depends on the available illuminating waveform and the ability of the receiver to recover weak modulated reflections in cluttered environments.

Sustainability also includes hardware lifetime, calibration burden, computational efficiency, and the number of deployed sensing nodes. Intelligent activation and cooperative node selection can reduce energy consumption by involving only those nodes that provide useful geometric or information gain for a given sensing task. The resulting optimization should balance sensing performance, communication quality, energy use, coverage, and infrastructure cost.

Figure~\ref{fig:section8_compare} summarizes the progression from baseline cellular sensing capabilities to broader 6G ISAC research directions. Overall, the evolution of ISAC should be evaluated not only through sensing resolution, estimation accuracy, or detection range, but also through its contribution to intelligent network control, distributed environmental awareness, resilience, energy efficiency, and service-level utility. The principal conceptual shift is from a network that primarily transports information through the physical environment to one that also observes, interprets, and responds to that environment as part of its normal operation.

\section{Conclusion}
This survey has examined the evolution of ISAC within 3GPP, from the Release~19 service and channel-model foundations established through TR~22.837 \cite{tr22837}, TS~22.137 \cite{ts22137}, and RP~234069 \cite{rp234069}, to the broader Release~20 activities addressing radio design, protocol procedures, and system architecture through TR~38.765 \cite{tr38765}, TR~23.700-14 \cite{tr2370014}, and TS~23.137 \cite{ts23137}. It has also discussed the wider 6G use cases, scenarios, and evaluation considerations described in TR~22.870 \cite{tr22870} and TR~38.914 \cite{tr38914}. By connecting SA1 service requirements, RAN1 physical-layer design, RAN2 and RAN3 system integration, and sensing-assisted communication, the survey provides a unified standards-centric perspective on cellular sensing. It also distinguishes standardized requirements and approved study foundations from study-stage assumptions and broader research directions.

Several challenges must be addressed to support scalable deployment. These include reproducible sensing-channel benchmarks, sensing-aware waveform and reference-signal design, practical self-interference suppression, multi-node coordination and data fusion, efficient measurement reporting and service exposure, and trustworthy governance of sensing information. More fundamentally, the development of ISAC depends not only on improvements in detection or estimation performance, but also on the coordinated evolution of service requirements, radio procedures, protocol mechanisms, network architecture, hardware capabilities, and security and privacy frameworks. ISAC ultimately represents a transition from communication networks that primarily transport information to networks that can also observe, interpret, and respond to their physical environment.

\bibliographystyle{IEEEtran}
\bibliography{references}

\end{document}